\documentclass[12pt,preprint]{aastex}

\begin{document}

\title{The Structure of the Star-forming Cluster RCW 38 }

\shorttitle{Structure of RCW 38}

\author{E. Winston\altaffilmark{1}, S. J. Wolk\altaffilmark{2} , T. L. Bourke\altaffilmark{2}, S.T. Megeath\altaffilmark{3}, R. Gutermuth\altaffilmark{4,5}, B. Spitzbart\altaffilmark{2}}

\altaffiltext{1}{ESA-ESTEC (SRE-SA), Keplerlaan 1, 2201 AZ Noordwijk ZH, The Netherlands. }
\email{ewinston@rssd.esa.int}
\altaffiltext{2}{Harvard Smithsonian Center for Astrophysics, 60 Garden St., Cambridge MA 02138, USA.} 
\altaffiltext{3}{Ritter Observatory, Dept. of Physics and Astronomy, University of Toledo, 2801 W. Bancroft Ave., Toledo, OH 43606, USA. }
\altaffiltext{4}{Five Colleges Astronomy Department, Smith College, Northampton, MA  01027}
\altaffiltext{5}{Department of Astronomy, University of Massachusetts, Amherst, MA  01003}

\begin{abstract}

We present a study of the structure of the high mass star-forming region RCW~38 and the spatial 
distribution of its young stellar population.   {\it Spitzer} IRAC photometry (3-8${\mu}$m) are 
combined with 2MASS near-IR data to identify young stellar objects by IR-excess emission 
from their circumstellar material.  {\it Chandra} X-ray data are used to identify class III pre-main sequence stars 
lacking circumstellar material.   We identify 624 YSOs: 23 class 0/I and 90 flat spectrum protostars, 
437 Class II stars, and 74 Class III stars.  We also identify 29 (27 new) O star candidates over the IRAC field. 
Seventy-two stars exhibit IR-variability, including seven class 0/I and 12 flat spectrum YSOs.   A further 177 
tentative candidates are identified by their location in the IRAC [3.6] vs. [3.6]-[5.8] cmd.  
We find strong evidence of subclustering in the region. Three subclusters were identified surrounding 
the central cluster,  with massive and variable stars in each subcluster.    The central region shows 
evidence of distinct spatial distributions of the protostars and pre-main sequence stars.     A previously 
detected IR cluster, DB2001\_Obj36, has been established as a subcluster of RCW 38. This 
suggests that star formation in RCW 38 occurs over a more extended area than previously thought.
The gas to dust ratio is examined using the X-ray derived hydrogen column density, $N_H$ and the 
$K$-band extinction, and found to be consistent with the diffuse ISM, in contrast with Serpens \& NGC1333. 
We posit that the high photoionising flux of massive stars in RCW~38 affects the agglomeration of the dust grains.

\end{abstract}


\today

\section{\bf Introduction}

The RCW 38 cluster is a region of high mass star formation, lying at a distance of 1.7~kpc from Earth.   
The region is one of the closest high mass star forming regions known to contain more than a thousand 
members \citep{wol}, after clusters such as the Orion Nebula Cluster \citep{mue}, Cep~OB3b \citep{get06}, and 
NGC~2244  \citep{wan}.     It was first identified in the $H\alpha$ survey of \citet{rcw} and the $HII$ 
survey of \citet{gum}.  The heart of the RCW~38 cluster is dominated by the O5.5 binary RCW~38 
IRS~2 \citep{fro,der}.  \citet{yama} obtained $^{13}CO$ NANTEN data of a large region surrounding 
RCW~38, showing two clouds, the eastern one coincident with IRS~2, while the second extends to the 
west.    Previous observations of  RCW~38  by \citet{der} using VLT/NACO identified 314 members from 
a total of 336 detections in the $J$, $H$, $K_s$-bands, in a field 0.5~pc squared surrounding IRS2.   
RCW~38 has previously been observed with {\it Chandra} as reported by \citet{wol}, where 360 X-ray 
sources from a total of 460 were positively associated with the star forming region.    An extensive 
overview of the previous studies of RCW~38 is to be found in the Handbook of Star Formation article 
on the cluster \citep{wolkhandbook}.  

Why are observations of young high mass star forming regions important to our understanding of star 
formation?  Current estimates place more than 60\% of star formation in clustered environments \citep{car,meg}.    
It is therefore important to perform multi-wavelength studies of such regions to understand the influence 
of the cluster environment in the process of star and planet formation.   One fundamental issue is the  
the spatial structure of clusters and the evolution of that structure.  The morphology of the molecular cloud 
as it undergoes collapse affects the subsequent star formation in the region.  By examining the spatial 
distribution of the young stars in a cluster it is possible to trace the underlying structure of the molecular 
cloud from which they form.    Such studies may differentiate between regions undergoing spherically 
symmetric collapse and those were the cloud is fragmenting to form distinct subclusters \citep{fei2008}.     
 
Regions of high mass star formation within $\sim$2kpc have been shown to exhibit varying morphologies:        
For example,  \citet{wan11} and \citet{wol11} have recently published studies of Trumpler~15 \& 16 in Carina, 
a massive star forming complex at a distance of 2.3kpc, and find very different structures. Tr~15 has a compact 
spherical structure with a continuous envelope of dispersed stars and mass segregation, Tr~16 has seven 
subclusters and no evidence of mass segregation.  The subclusters in Tr~16 are contiguous; they contain 
massive stars but none are located within 0.2~pc of the cluster center.   Both regions have very low 
circumstellar disk fractions ($<$10\%) consistent with ages of 3-10Myr.  
NGC~2244 in the Rosette Nebula lies at a distance of 1.4~kpc and was recently studied by \citet{wan}.  
It has as an annular morphology and a central cluster of O-stars, with a larger scale spherical structure.  
There is no evidence of subclustering and a low disk fraction ($\sim$6\%) though its age is estimated at 2~Myr.  
The RCW~108 region, at a distance of 1.3~kpc, is a nearby example of triggered star formation \citep{com}, 
with evidence of mass segregation and its YSO population still embedded in its molecular cloud \citep{wol2}.
In RCW~38, where the disk fraction is consistent with an age of 1Myr, we will present evidence of four 
separate subclusters surrounded by a dispersed population, with evidence of mass segregation and possible 
triggering of a second generation of low mass stars surrounding IRS~2.  
   
Evolution of dust grains in the ISM and within molecular clouds can have important 
implications for the development of circumstellar disks and hence the formation of planets 
around young stars.    Of specific importance is the study of the effect of massive stars in 
densely populated star-forming regions, to examine the effects of ionising radiation on the 
evolution of disks around lower mass stars, and to examine the star formation efficiency 
where much of the gas and dust are processed by stellar radiation fields. 
The gas to dust ratio can be examined through the measured dust extinction in the IR and 
the X-ray calculated hydrogen column density.   In previous studies of less massive clusters, 
the authors found possible evidence for grain growth in Serpens North and NGC~1333  
\citep{winston,win10}.  Higher mass regions, such as RCW~108 and LkH$\alpha$~101 have 
been found to be similar to the ISM  \citep{wol2,wol3}. \citet{wol} previously found a value of 
$N_H = 2\times10^{21}~A_V$ for this relation in RCW~38.    The implication that the presence 
of massive stars in the vicinity impedes grain growth will here be re-examined making use of 
the improved IRAC evolutionary classification of the YSOs.    

Observations with the {\it Spitzer} and {\it Chandra} Space Telescopes provide 
the means to identify  young stellar objects from the protostellar through the diskless pre-main 
sequence phase. The IR data further allows for the categorization of the 
objects into the canonical evolutionary classes (class 0/I, II and III \citep{lad84}).  
The X-ray data of \citet{wol} are utilized here in conjunction with the {\it Spitzer} IRAC data to identify young 
diskless members of the cluster.
Information on the distribution of star formation sites, the dispersal after formation of the YSOs, 
and the dynamical state of the complex as a whole can be found by examining the number and 
distribution of the YSOs.  Such studies provide a better understanding of the physical processes 
that govern the collapse and fragmentation of molecular clouds, the effect of massive stars on the 
evolution of their less massive neighbors, and of the possible future dispersal of the cluster after 
the natal gas has been dispersed.     
Clusters also serve as laboratories for the evolution of young stellar objects, and can be used to 
study the  evolution of disks around young stars and the evolution of hot X-ray emitting plasma 
commonly found around such stars (see for example: \citet{gut09,hai,pre3,her,winston,wol3}).  

In this paper we present the {\it Spitzer} IRAC observations of the extended RCW~38 
cluster, identifying 624 YSOs in the region. We first present a description of the 
{\it Spitzer} data and its reduction. 
Second, we discuss the techniques used in the identification of the cluster members, 
and the evolutionary classification of the young stars. 
We then discuss the X-ray characteristics of the stars as compared to the infrared characteristics, 
the temporal photometric variability of the young stars, the selection of O-star candidates in the 
region,  and lastly the spatial distribution of the cluster members is presented.

\section{\bf Observations and Data Reduction}\label{obsdata} 

\begin{figure}
\epsscale{1}
\plotone{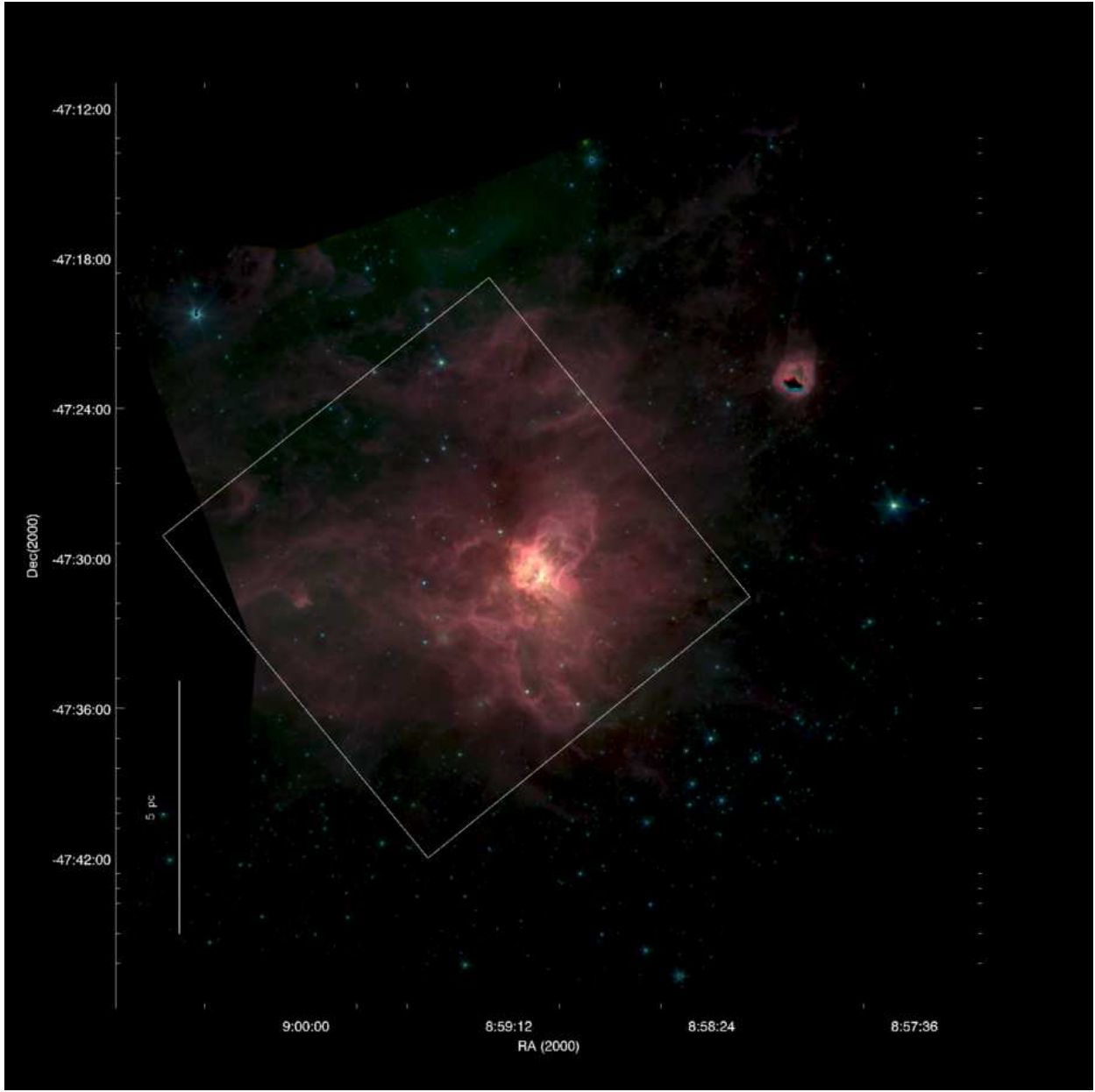}
\caption[]{ The RCW 38 region observed with IRAC on $Spitzer$.  The plot shows 
a three-band false color image of the cluster, where the mosaic at each wavelength 
was created from the four epochs of data combined using the $Montage$ mosaicing 
software. The field shows the overlap region of the four IRAC bands.   
Blue is 3.6${\mu}m$, green is 4.5${\mu}m$, and red is 8.0${\mu}m$.  The reddish 
hue at 8.0${\mu}m$ is due mainly to diffuse PAH emission.   Emission from shocked 
hydrogen is visible in green. The outline of the Chandra ACIS-I field of view is 
overlaid as a white square. }
\label{field}
\end{figure}

\begin{figure}
\epsscale{1}
\plotone{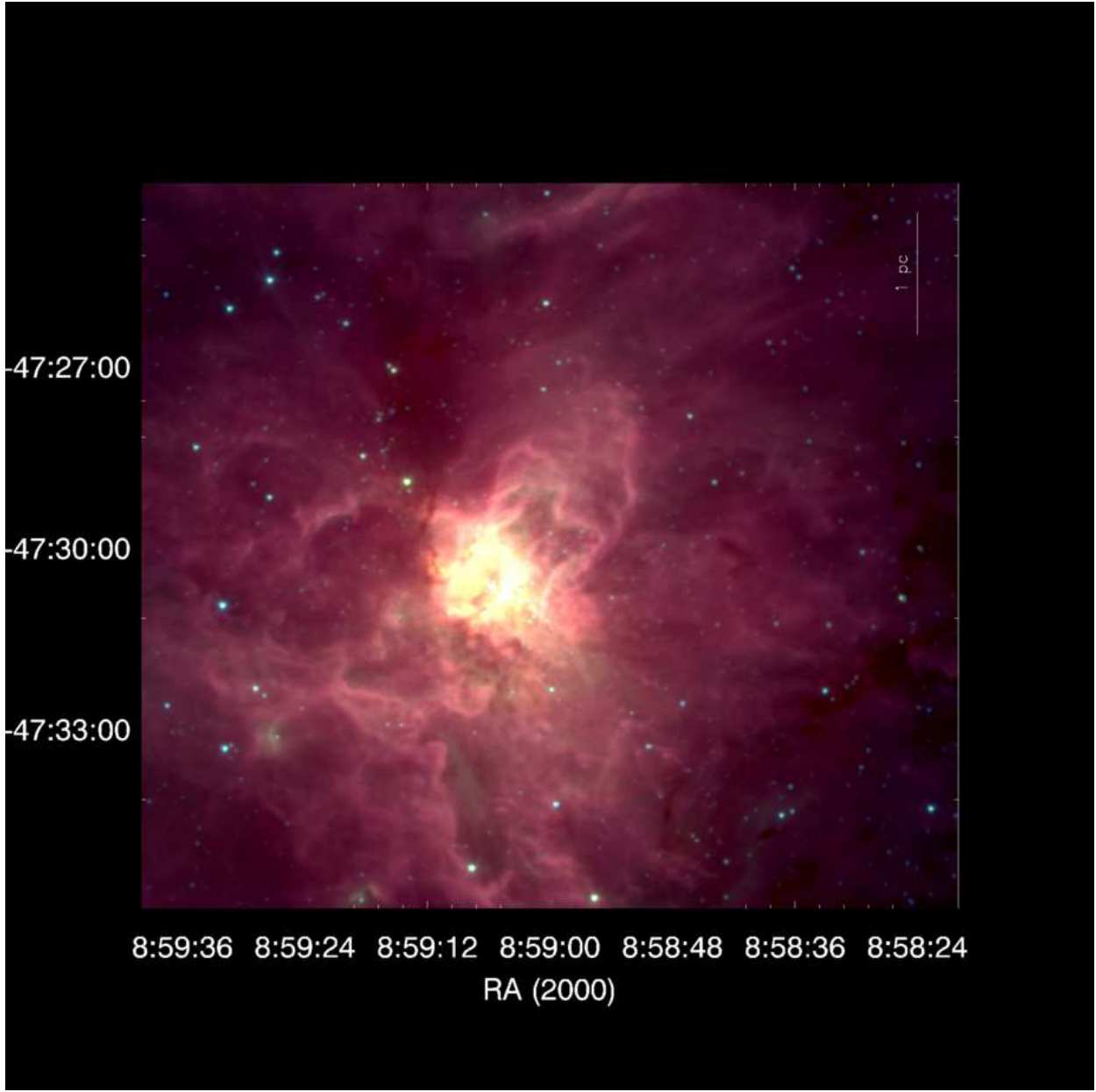}
\caption[]{ The RCW 38 central core, surrounding IRS 2. The plot shows 
a three-band false color image of the cluster using IRAC on $Spitzer$. 
Blue is 3.6${\mu}m$, green is 4.5${\mu}m$, and red is 8.0${\mu}m$.
The reddish hue at 8.0${\mu}m$ is due mainly to diffuse PAH emission. 
Emission from shocked hydrogen and $Br_{\alpha}$ is visible in green. }
\label{core}
\end{figure}

We have obtained observations of the RCW 38 region with the 
{\it Spitzer} InfraRed Array Camera (IRAC; \citet{faz}) in four 
wavelength bands: 3.6, 4.5, 5.8, and 8.0~$\mu$m at four different epochs.  
Fig.~\ref{field} and Fig.~\ref{core} show IRAC three band false-color images of 
the combined field of view of the four epochs and the central core of RCW 38.  
We also obtain data with  at 24~$\mu$m from the Multiband Imaging Photometer 
for {\it Spitzer} (MIPS; \citet{rie}) at two separate epochs.  
This photometry was supplemented by $J$, $H$ and $K$-band photometry 
from the 2MASS point source catalogue \citep{skr}.   
In the following analysis, only data with uncertainties $\le 0.2$~magnitudes  
were used.   Below, we describe the observations, image reduction and photometry 
for the {\it Spitzer} data.

\subsection{Mid-IR:  $Spitzer$ IRAC \label{irac}}

The {\it Spitzer} Space Telescope has observed the RCW~38 cluster  
with the IRAC instrument at four different epochs as part of PID~20127.  
This observing strategy purposed to fully cover the $\sim$30$'$$ \times$30$'$ 
field while recovering the inner $\sim$15$'$$ \times$15$'$ with shorter exposures 
to overcome the high background of the central core.    
The first epoch, Epoch1, was taken on 2006-06-05 and utilized cluster mode to map 
the outer field in a spiral pattern with each pointing offset by $\sim$270$\arcsec$, with 
12 second frames and 5 dithers per position with the cycling dither turned on.  
The map ends before the interior region of $\sim$6$'$$ \times$6$'$ was observed to 
avoid the build-up of short-term latents on the image.  
The second epoch, Epoch 2, was taken directly following Epoch 1 on 2006-06-05 
but observed a smaller $\sim$15$'$$ \times$15$'$ central field of RCW~38. 
An integration time of 2 seconds with 1 dither per position with the cycling dither turned on 
was used to avoid saturation.  
The third epoch, Epoch 3, was observed on 2007-02-21 and followed the same spiral pattern
over the outer field of $\sim$30$'$$ \times$30$'$ surrounding the RCW~38 core, again 
ending before the core was observed.  
The last epoch, Epoch 4, directly followed Epoch 3 and was taken on 2007-02-21 
and also observed a $\sim$15$'$$ \times$15$'$ IRAC overlap region of the RCW~38 core.   
Table~\ref{epochs} lists the integration times and coverage of the four epochs.

The analysis used the basic calibrated data (BCD) FITS images from the
S14.0.0 pipeline of the {\it Spitzer} Science Center for Epochs 1 \& 2 and the 
S15.3.0 pipeline for Epochs 3 \& 4.   
These data were combined into mosaics using $WCSmosaic$, which also removed 
cosmic ray and other artifacts.  
Custom IDL routines, part of the $clustergrinder$ package \citep{gut09} were used to clean 
the mosaics of bright source artifacts.  
The pixel scale of the mosaics is 1.2$"$/pixel. 
The calibration uncertainty across the IRAC bands is estimated at $0.02$~mag \citep{rea}.
In addition, there are $\sim$5\% position dependent variations in the calibration 
of point sources in the flat-fielded BCD data; these have not been corrected for 
in our data.  Furthermore, these systematic uncertainties have not been added 
into the photometric uncertainties reported in this paper.
The noise in a $8 \times 8$ pixel region around the source was calculated 
to assess the combined instrumental noise, shot
noise and noise from the spatially varying extended nebulosity. 
Point sources were found using $Photvis$, based on a 
heavily modified version of the DAOfind program in the IDLPHOT package \citep{lan}.  
Point sources with peak values more than 5 $\sigma$ above the background 
were considered candidate detections.   After identifying the point sources, 
aperture photometry was obtained using the $aper.pro$ program in the IDLphot package.  
An aperture radius of 2 pixels was used, and a sky annulus from 2 to 6 pixels was used 
to measure the contribution from extended emission in the aperture with $mmm.pro$, 
also in the IDLphot package.  
The zero points for these apertures were:  17.0879, 16.5843, 15.9790, and 15.1945~mag for the 
3.6, 4.5, 5.8, 8.0~${\mu}m$ bands respectively, in the native BCD image units 
of MJy sr$^{-1}$.  

In Epoch 1, there were 4195 detections in the 3.6${\mu}m$ band with uncertainties less than 0.2~mag, 
and 3346, 1536, 653, in the 4.5, 5.8, and 8.0${\mu}m$ bands, respectively, with 302 detections in all 
four IRAC bands and 816 detections in the three shortest IRAC bands.  
In Epoch 3, there were 4137 detections in the 3.6${\mu}m$ band, and 3941, 1596, 893, 
in the 4.5, 5.8, and 8.0${\mu}m$ bands, respectively, with uncertainties less than 0.2~mag, 
with 415 detections in all four IRAC bands and 1,039 in the three shortest IRAC bands.  
In the smaller deeper field Epoch 2, there were 2106 detections in the 3.6${\mu}m$ band with 
uncertainties less than 0.2~mag, and 1605, 713, 205, in the 4.5, 5.8, and 8.0${\mu}m$ bands, 
respectively, with 108 detections in all four IRAC bands and 329 in the three shortest bands.  
In Epoch 4, there were 1775 detections in the 3.6${\mu}m$ band with uncertainties less than 
0.2~mag, and 1844, 454, 299, in the 4.5, 5.8, and 8.0${\mu}m$ bands, respectively, with 116 
detections in all four IRAC bands and 311 in the three shortest bands.  
The sensitivity of the 8.0~${\mu}m$ band is the limiting factor in four band detections: it suffers 
from lower photospheric fluxes, higher background emission, and the presence of bright, 
structured nebulosity across much of the image. 
The arrangement of the detectors ensures that only a fraction of the sources can have 4-band 
detections in any single epoch (c.f Sec.\ref{cat_merge})
The detection threshold in the 3.6~${\mu}m$ band was $\sim$17 magnitudes - just at the 
hydrogen burning limit for a 1~Myr old star at the distance of RCW~38 ($\sim$16.5~mag 
at 3.6~$\mu$m \citep{bar}).

\subsection{MIPS:}

We observed the RCW~38 cluster on two occasions with the Multiband
Imaging Photometer for {\it Spitzer} (MIPS; Rieke et al. 2004).  The observations 
were taken as part of PID~20127. 
The first epoch was observed on 2006-03-03 and had an aim point of $8^{hr}59^{m}5.78^{s}$, 
$-47^{d}30^{m}43.02^{s}$ (J2000).  The second epoch was observed on 2006-06-18 and 
had an aim point of $8^{hr}59^{m}5.49^{s}$, $-47^{d}30^{m}41.14^{s}$ (J2000).  
In both epochs, the medium scan rate was used and the total map sizes were 
approximately 0.5 x 0.5 degrees including the overscan region. 
The typical integrated exposure time per pixel is about
2.62 seconds at 24~$\mu$m.  The MIPS images were processed using the MIPS 
instrument team Data Analysis Tool, which calibrates the data and applies a 
distortion correction to each individual exposure before combining into a final 
mosaic \citep{gor}.  The data were further processed using various median
filters to remove saturation effects and column-dependent background
structure.  The resulting mosaics have a pixel size of $1.25''$ at 24~$\mu$m.  
The 24~$\mu$m images suffered from saturation of the central core region and 
strong extended emission over the field; it was not possible to extract reliable 
photometry for any objects, instead the 24~$\mu$m mosaic was used to search 
for deeply embedded protostars surrounding the core (cf. Sec.~\ref{deo}).

\subsection{Chandra:}

This paper utilizes the previous $Chandra$ observations of RCW~38, first  presented in \citet{wol} 
and now available on the ANCHORS archive (AN archive of Chandra Observations of Regions of 
Star formation\footnote{http://cxc.harvard.edu/ANCHORS/}).  
The observations were reprocessed and an updated source detection procedure was applied 
for the archive, and this reprocessed dataset is used here.   
Originally, the data were processed in 2001 through the CXCDS $ver.$6.12 pipeline.   
A  wavelet based source detection algorithm, originally developed for ROSAT, known as 
PWDetect \citep{dam}  was then used to identify sources across the entire ACIS-I array.   
The array was 2x2 binned and the  threshold significance was set to detect sources between 
4 and 5.31 equivalent Gaussian sigmas; the data are searched on scales of 0.5 to 16''. 
With these settings, a false detection rate of $< 1\%$ is expected. A total of 460 sources were found. 
The 2x2 binning, employed due to memory constraints means the the image is 
under-sampled with respect to the maximum resolution of ACIS. Hence, some close 
sources, on-axis, can be conflated. 
For the new source extraction, a three-pass method was used, with the CIAO 3.4 Wavedetect tool 
\citep{fre}.  The three-pass method searches for sources in images binned by 1, 2 
and 4 pixels, respectively, using wavelet scales of 2, 4, 8, 16 and 32 pixels on each image.
Each image has a size of 2048x2048 pixels. This method was designed to optimally take 
advantage of additional memory more commonly available in 2007.  The source 
limits were again set to limit false detections to about $1\%$, however Wavdetect 
has more robust error analysis and tests indicated that this was slightly more sensitive 
than PWDetect at the nominal $1\%$ false detection probability.  The results of the three 
passes are combined into a single list of  536 sources.  This includes 18 sources 
detected on the S array which was not considered to by \citet{wol}.  By comparison 
the Chandra Source Catalog \citep{eva} detects 410 sources in the field with a 
minimum flux threshold of about 10 net counts on axis, which is significantly higher 
than our equivalent limit.   
Table~\ref{newxrays} lists the identifiers, coordinates and properties of the 536 sources 
identified in this reprocessed dataset.  
The table provides the {\it Chandra} identifier, source locations, the raw and net number counts (net counts are 
background subtracted and aperture corrected), plasma temperature ($kT$), hydrogen column density ($N_H$),  
absorbed and unabsorbed X-ray flux ($F_X$), and flaring statistics.

\subsection{Catalog Merging}\label{cat_merge}

For each epoch of {\it Spitzer} data, the photometry of the four IRAC bands were merged with 
the 2MASS $J$, $H$, and $K$-band and {\it Chandra} X-ray data to form an initial catalogue 
using custom IDL routines. 
Sources observed in different wavelength bands, which were located within $1''$ of each other 
were merged in the catalogue; if while comparing two bands, multiple sources in one of the 
bands satisfied this criteria, the closest was chosen.
The {\it Chandra} X-ray data were taken from \citet{wol}, where a full description of the data 
reduction process was presented.   The observation,  {\it Chandra} obsid 2556, had a total 
exposure time of 97~ks, with 536 sources detected in the {\it Chandra} field.  
These catalogues were used in the identification of the YSOs, c.f. Sec.\ref{idysos}.

The fields of view (FOVs) at each IRAC epoch, and of the different instruments did not 
cover the same extent of the cluster, with the IRAC epochs 2 \& 4 limited to a  
15$'$ $\times$ 15$'$ field in the central cluster.  The two larger fields, epochs 1 \& 3, cover a 
30$'$ $\times$ 30$'$ FOV of the cluster, with a $\sim$6$'$$ \times$6$'$ gap in coverage to avoid 
saturation in the cluster core.  Furthermore, the IRAC detectors are split into two groups, 
channels 1 \& 3 (3.6 \& 5.8~${\mu}m$) and channels 2 \& 4 (4.5 \& 8.0~${\mu}m$), whose 
FOVs are offset from one another by $6.5'$.   The {\it Chandra} field is the smallest overall, 
covering a region 17$'$ $\times$ 17$'$, centered on IRS 2.  The X-ray sample is thus limited 
to the central region of the cluster, and data are not available for all IR sources in the catalogue.   
The 2MASS data are not spatially constrained.   
A final photometric catalogue was then created by matching the sources detected in the four 
initial catalogues to within $1"$.  This catalogue was used in the variability study, c.f. Sec.\ref{variability}.

\section{\bf Identification of YSOs}\label{idysos}  

The RCW 38 field contains 26,873 sources with a detection in at least one IR band, 
at one epoch, across the entire IR field. In the IRAC overlap field, there were 13,950 
sources with a detection in at least one IR band.  
In order to identify the possible young stellar objects (YSOs) in the four epochs, two 
selection techniques were applied to the data: selection of stars with IR excesses on 
IR color-color diagrams, and identification of X-ray luminous YSOs by comparison 
of X-ray sources with IR detections.  
These methods were applied to each of the four IRAC epochs separately in the IR field, 
after which, the four catalogues of candidate YSOs were merged into one master list of 
candidate members of RCW 38.  This list was then trimmed of those candidates not 
appearing in the IRAC overlap field of at least one epoch.  

In each epoch, the field was limited to that region observed by all four IRAC bands: the 
overlap IRAC field.  By examining this restricted field, we insure that each individual source 
has the possibility of being detected at all four IRAC wavelengths.
Each of the four IRAC epochs was examined individually from the others, but each was 2
combined with the same X-ray and near-IR observations.

\subsection{Infrared Excess Emission}

\begin{figure}
\epsscale{1.}
\plotone{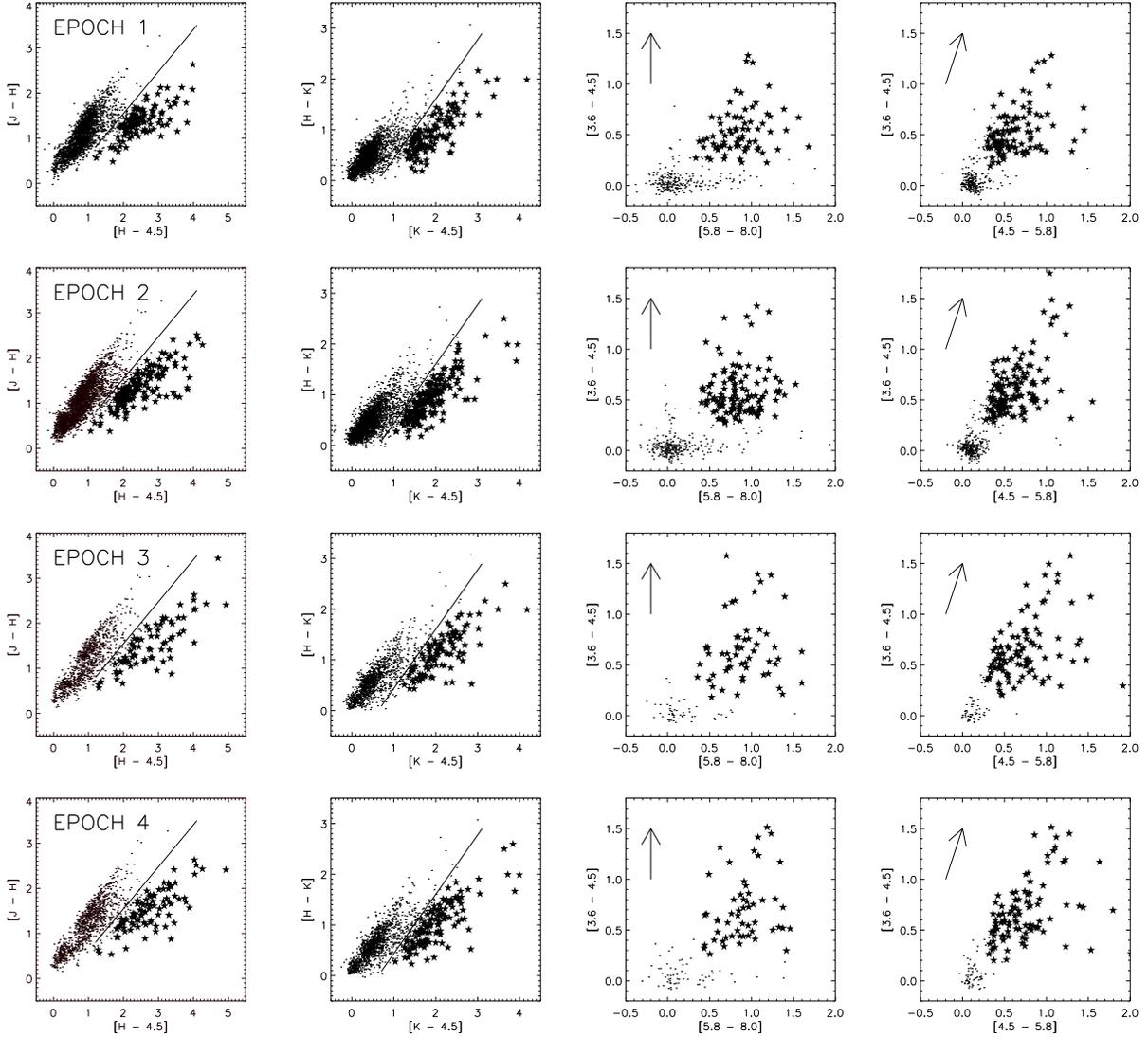}
\caption[]{ Four color-color diagrams used to identify YSOs, top to bottom: Epochs 1 - 4. 
{\it Left plots:} $J - H$ vs. $H - [4.5]$ diagrams. The solid line shows the reddening vector for these 
wavelengths; objects with $H - [4.5]$ greater than 1$\sigma$ below the reddening vector are 
considered to have an excess.   The black dots indicate those objects classified as having no 
excess in this diagram, while the stars indicate those sources with an excess.  
{\it Centre Left plots:} $H - K$ vs. $K - [4.5]$ diagrams. Similar to the right plots, with stars indicating 
those sources with excess in their $K - [4.5]$ colors.  
{\it Centre Right plots:}  IRAC color-color diagrams, $[3.6] - [4.5]$ vs. $[5.8] - [8.0]$.  The locus for field 
stars lies on the origin; the spread in the $[5.8] - [8.0]$ colors is in part due to contamination from 
nebulosity and contamination from star-forming galaxies with strong PAH emission in the 5.8 and 
8.0${\mu}m$ bands.  A reddening vector of $A_K = 5$  is shown.       
{\it Right plots:}  IRAC color-color diagrams, $[3.6] - [4.5]$ vs. $[4.5] - [5.8]$.  The locus for field 
stars lies near the origin; the $[4.5] - [5.8]$ color is in part due to contamination from 
nebulosity, such contaminants are subsequently filtered.  A reddening vector of $A_K = 5$  is shown.       
  }
\label{figccds}
\end{figure}

Young stellar objects can be identified by their excess emission at IR
wavelengths. This emission arises from reprocessed stellar radiation
in the dusty material of their natal envelopes or circumstellar disks.
The infrared identification of YSOs is carried out by identifying
sources that possess colors indicative of IR excess and distinguishing
them from reddened and/or cool stars \citep{meg,all,gut1}.  
The main limitation of this method is the contamination from extragalactic 
sources such as PAH-rich star-forming galaxies and AGN, and galactic 
sources such as AGB stars; both of these groups have colors similar to 
young stellar objects.  Four color-color diagrams 
were used to determine the color excess of the sources: 
two IRAC diagrams: [3.6] - [4.5] vs. [5.8] - [8.0] and [3.6] - [4.5] vs. [4.5] - [5.8], 
and two IRAC-2MASS diagrams: $J - H$ vs. $H - [4.5]$ and $H - K$ vs. $K - [4.5]$.  
Fig.~\ref{figccds} shows the four diagrams for each IRAC epoch (Ep.1-4; note that the 
2MASS photometry is the same throughout).  In the following analysis we 
required photometry to have uncertainties $< 0.2$~mags in all bands 
used for a {\it particular} color-color diagram.  
The numbers of sources for each diagram which satisfy this criteria are, respectively, 
in Epoch 1:  302, 816, 2748, 3007; in Epoch 2:  108, 329, 1084, 1342; in Epoch 3: 
415, 1039, 3068, 3478; in Epoch 4:  116, 311, 1195, 1534.

\subsubsection{IRAC Color-Color Diagram}

The main effort involved in identifying  bone fide IR excess sources is 
distinguishing between reddened or cool stars and those with excess 
emission arising from circumstellar material.  
A reddening law in the IRAC bands was determined by \citet{fla}, which 
shows the [5.8] - [8.0] color to be particularly insensitive to reddening and 
stellar temperature, and is thus a reliable measure of excess emission due to
dust.  Sources with a color $> 1 \sigma$ beyond $[5.8] - [8.0] > 0.2$
mag are likely to possess excesses and not to be reddened or cool
stars.  Extragalactic sources such as PAH rich star-forming galaxies 
and AGN will also satisfy this criteria \citep{ste}.
Sources with a color more than $1~\sigma$ below $[3.6] - [4.5] < 0.1$ 
were considered galaxies and were removed. 
AGB stars exhibit colors similar to reddened or cool stars over the IRAC 
bands \citep{blum} and are filtered out with the field stars.   

\citet{gut09} have recently developed a method for substantially 
reducing extragalactic contamination built on the Bootes Shallow Survey 
data \citep{eis} and the {\it Spitzer cores to disks} legacy program methods \citep{jor,har}.   
The galaxies are eliminated either by their colors, which unlike YSOs are 
often dominated by PAH features in the 5.8 and 8.0~${\mu}m$ bands, or 
by their faintness.  It should be noted that very faint or embedded protostellar 
sources may be erroneously filtered by this method. 
A total of 69 candidate YSOs were selected from the IRAC color-color diagram
in Epoch 1, with 56 in Epoch 2, 119 in Epoch 3, and 54 in Epoch 4.  Of these 17 
were identified as contaminating sources and removed from the Epoch 1 catalogue, 
8 from Epoch 2, 11 from Epoch 3, and 9 from Epoch 4.

\subsubsection{IRAC-2MASS Color-Color Diagrams}

The shorter wavelength 3.6 and 4.5~${\mu}m$ IRAC bands are far more
sensitive to stellar photospheres than the longer wavelength 5.8 and
8.0~${\mu}m$ bands which suffer from higher backgrounds; e.g. in Epoch 1 
where the shorter wavelength bands have 4195 and 3346 detections  with 
$\sigma \le 0.2$, respectively, the longer wavelength bands have
1536 and 653 detections. Hence many detections cannot be
characterized on the IRAC color-color diagram.  For this reason, it
is important to develop methods to identify sources with infrared
excesses that rely only on the shorter wavelength bands.  

We combine the near-IR data from the 2MASS point source catalog with the 4.5~${\mu}m$ band 
photometry \citep{skr}.  This provides the ability to detect dust around objects which
are too faint for detection in the 5.8 and 8.0~${\mu}m$ bands.  In
particular, we concentrate on the $J-H$ vs. $H-[4.5]$ diagram and the
$H-K$ vs $K-[4.5]$ diagram  \citep{winston}.  These diagrams take advantage of the
sensitivity of {\it Spitzer} in the 4.5~$\mu$m band and the stronger infrared 
excess emission at 4.5~$\mu$m compared to that at shorter wavelength \citep{gut2}.   
For highly reddened sources which are not detected in the $J$-band, the 
$H-K$ vs. $K-[4.5]$ diagram can be used.
It should be noted that the IRAC 4.5~${\mu}m$ data can detect
sources too faint or reddened to have been detected by 2MASS.   
AGN are typically fainter than stellar YSOs found in star-forming
regions; all candidates with 3.6~${\mu}m$ magnitudes fainter than 16~mag that 
lacked the required bands to be placed on the \citet{gut09} diagrams were removed.    
Utilizing the $J-H$ vs. $H-[4.5]$ and $H-K$ vs. $K-[4.5]$ diagrams,  in 
Epoch 1, 109 and 120 sources were selected as having an IR excess, 
72 and 99 in Epoch 2, 160 and 180 in Epoch 3, and 86 and 118 in Epoch 4,
respectively.

\subsubsection{IRAC 3-Band Color-Color Diagram}

Due to the limiting factor of requiring an 8~$\mu$m detection to place a source on the 
IRAC CCD, we also utilized the IRAC 3-band CCD [3.6]-[4.5] vs. [4.5]-[5.8] to identify 
sources with excess emission but with photometry only in the 3.6-5.8$\mu$m bands.  
The criteria for young excess sources on this diagram are presented in \citet{gut09}. 
Contamination from shocked emission and PAH contamination is dealt with by the 
selection criteria, however some AGN interlopers may remain. Sources with 
3.6~${\mu}m$ magnitudes fainter than 16~mag were removed as possible AGN. 
In Epoch 1, 101 sources were identified as having an excess using this method, of 
which 24 were not previously known from the other IR diagrams. In Epoch 2, 145 sources 
were identified, of which 27 were new.  In both Epochs 3 \& 4, 101 objects were identified, 
with 26 and 25 new sources, respectively.  Combining the four epochs, a total of 33 
additional YSOs were identified using this method.   

\begin{figure}
\epsscale{1.}
\plotone{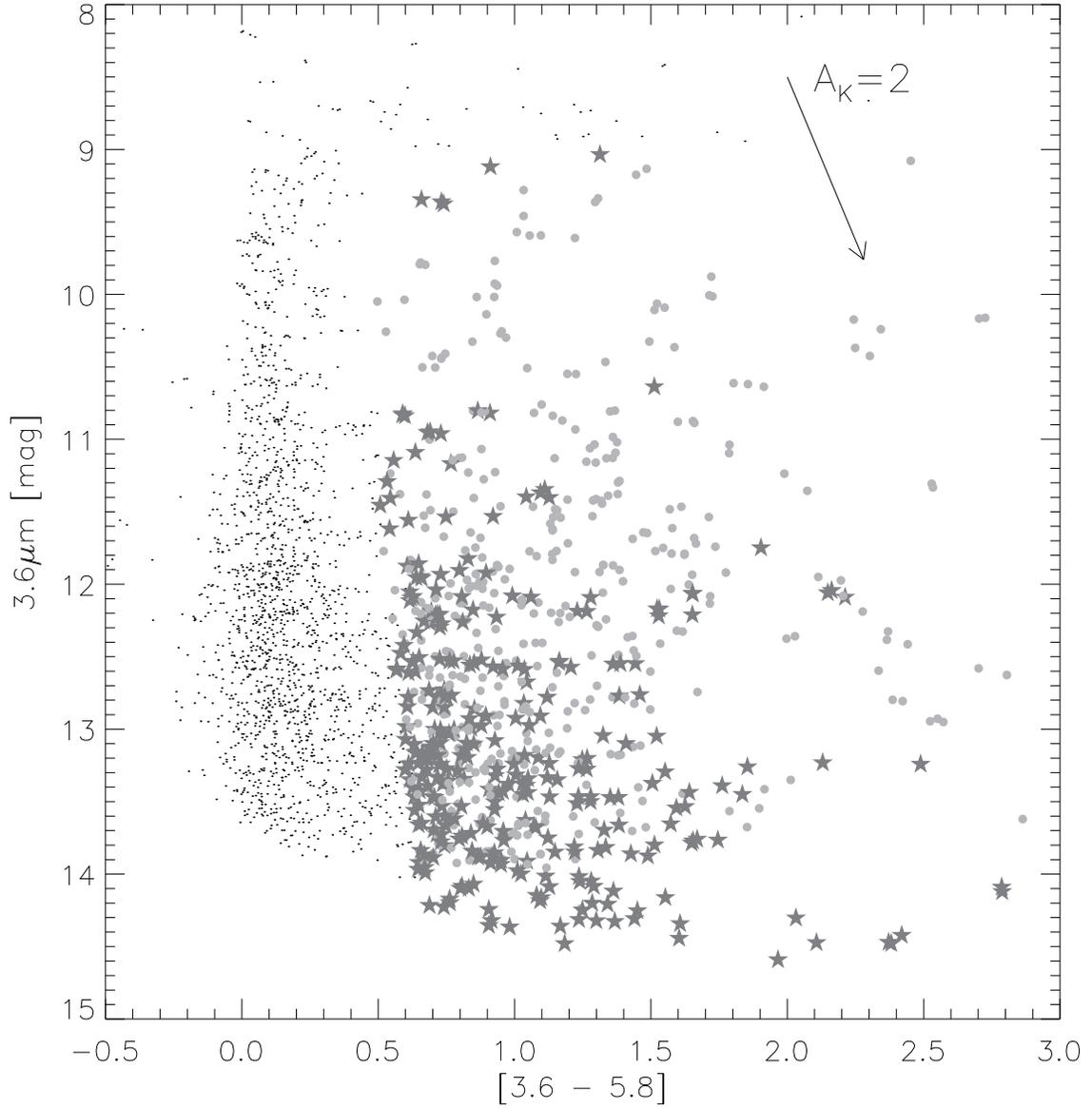}
\caption[]{ $[3.6]$ vs. $[3.6] - [5.8]$ color-magnitude diagram used to identify candidate YSOs.  
The field stars are shown as black dots, the previously identified YSOs as grey circles, and the 
newly selected candidates as dark grey stars.  Sources with $8.0 \le m_{3.6} \le 14.0$ and a 
color more than $1~\sigma$ greater than $[3.6 - 5.8] > 0.4$ were selected.  A further 177 
candidate YSOs were identified.           
  }
\label{fig3cmd}
\end{figure}

In addition to this method, the color-magnitude diagram of [3.6] vs. [3.6-5.8] was employed 
to identify possible cluster members not identified on the previous diagrams due to weak 
or no excess emission from the disk or a lack of photometry at 8.0$\mu$m.   
Figure~\ref{fig3cmd} shows the  [3.6] vs. [3.6-5.8] cmd, with field stars shown as black 
dots, the previously identified YSOs as grey circles, and the newly selected candidates as 
dark grey stars.  
Those sources with $8.0 \le m_{3.6} \le 14.0$ and a color more than $1~\sigma$ 
greater than $[3.6 - 5.8] > 0.4$ were selected.  This region delimited the color-magnitude 
space of the previously identified YSOs on this diagram. We identified a further 177 candidates 
with the appropriate colors.  Given that contamination cannot be identified using the 
contamination removal methods outlined in \citet{gut09}, we consider these to be tentative 
candidate members only and do not include them in the following discussions, c.f. Sec.~\ref{discussion}.
The coordinates and photometry of the 177 candidates are presented in Table~\ref{tablephotcandy}.

\subsubsection{Completeness}

\begin{figure}
\epsscale{1}
\plotone{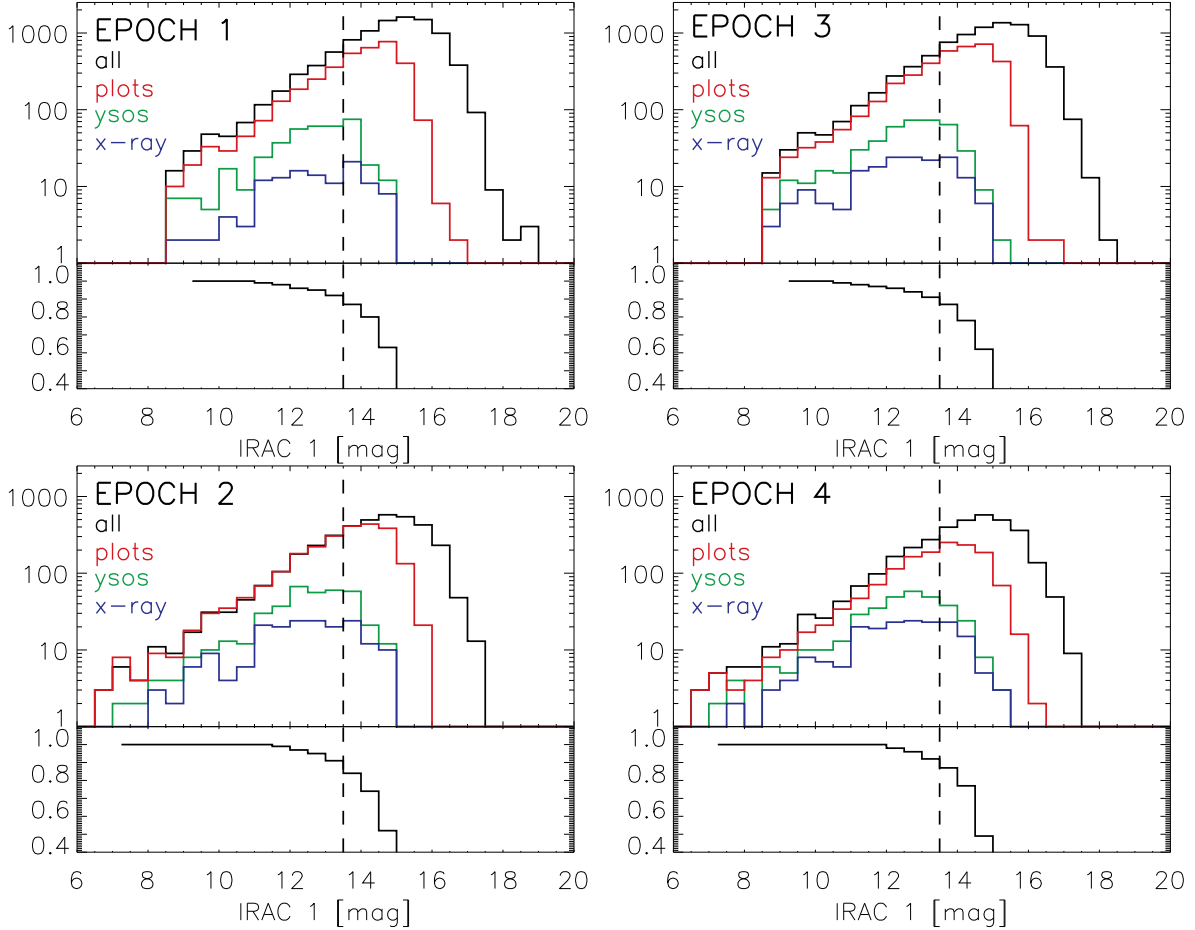}
\caption[]{
{\it Above:} Histograms of all photometry with a 3.6${\mu}m$ detection in the
IRAC four band overlap region, by magnitude, as a solid black line. 
The red line shows those sources with photometry in the multiple bands required  
to place them on the color-color diagrams used to select IR excess sources.  
The green line shows the final YSO catalog. The blue line plots the subset of YSOs 
in the final catalog with a $Chandra$ detection; 
note that the X-ray observations covered a smaller field than the IR observations.   
{\it Below:} Fraction of artificial stars recovered as a function of magnitude for the 
3.6${\mu}m$ photometry. The dashed black line indicates 90\% completeness.  
 }
\label{complete}
\end{figure}

The RCW 38 cluster lies at a distance of 1.7~kpc and exhibits a highly variable background 
across the extent of the cluster.  Completeness limits for the {\it Spitzer} photometry were 
assessed using a method of inserting artificial stars into the mosaics and then employing 
our detection algorithms to identify them.  Estimates for the 90\% completion limits for all 
epochs in each of the IRAC bands were 13.5, 13.0, 12.0, 10.5~mag, respectively.  These 
limits correspond to an approximate stellar mass of 0.5~$M_{\sun}$ at 3.6${\mu}m$ and an 
age of 1Myr \citep{bar}. 
Fig.\ref{complete} shows the magnitude histograms of the 3.6~${\mu}m$ photometry at each 
of the four epochs; all detected sources with uncertainties $\le0.2$ are shown by the upper 
black curve.   The red histograms show the sources with detections in bands required to place 
them on at least one of the IR excess diagrams.  The sample of identified YSOs is shown by 
the green histogram. The blue histogram shows the subsample of identified YSOs that possess 
an X-ray detection.  

The Hydrogen-burning limit, at 1.7~kpc, is $\sim$17 mag  at 3.6${\mu}m$ and an 
age of 1Myr \citep{bar}.  Though the photometry is sensitive to this magnitude,  at this distance 
source crowding and the bright, highly varying background become serious issues and, as can 
be seen from the histogram of identified YSOs, our identification threshold is closer to 15~mag 
at 3.6${\mu}m$, or $\sim$0.15~$M_{\sun}$ at 1Myr \citep{bar}.     
When considering objects that can be placed on the CCDs we are limited to $\sim$13.5 mag at 
3.6$\mu$m (or 0.5~$M_{\sun}$).   The ratio of the number of sources with $m_{3.6} \le 13.5$~mag 
that can be placed on the color-color diagrams to the total number of sources with 
$m_{3.6} \le 13.5$~mag (and correcting this number for completeness by dividing by the fraction 
of artificial stars recovered in each magnitude bin) is $1778/1883$ or 94\% for Epoch 1, $993/1094$ 
or 90\% for Epoch 2, $1822/1966$ or 92\% for Epoch 3, $867/975$ or 89\% for Epoch 4.   These 
fractions represent upper limits on the completeness of the sample, as the number of contaminating 
field stars is rising with increasing magnitude, while the number of YSOs does not show an 
equivalent rise.   

The \citet{der} study demonstrates the loss of detections towards the RCW~38 nebula in the IRAC 
data.    \citet{der} identify 483 sources with uncertainty $<$0.2 in at least one near-IR band in the 
central 0.5~pc surrounding IRS~2; in comparison, we detect 31 objects in at least one IRAC band 
with uncertainty $<$0.2 in the same region.  The measured stellar surface density in \citet{der} is 
ten times that found with {\it Spitzer}  due to the higher angular resolution available with the VLT 
and the reduced confusion with nebular emission achieved toward the RCW~38 nebula in the 
near-IR.   

A more conservative estimate for the 90\% completeness of the pre-main sequence membership 
of the cluster (those with IR-excess emission) can be made from the green histograms of 
Fig.~\ref{complete} as 12~mag at 3.6$\mu$m, or $\sim$2~$M_{\sun}$ at 1~Myr \citep{sie}.  This 
estimate does not account for extinction, which will move completeness to even higher masses.  
Further, completeness varies across the field due to the varying background emission, being  
particularly low in the central region surrounding IRS~2.   Since the fraction of sources with 
IR-excess varies with mass and age \citep{lad,her}, the completeness with respect to all pre-main 
sequence stars cannot be estimated with the available data.

\subsection{X-ray Luminous Objects}  

Young stellar objects possess elevated levels of X-ray
emission, with luminosities, $L_X$, of  $\frac{ L_{Xbol} }{ L_{X\sun} } \sim 10^{3.5} $ that can be used
to distinguish them from foreground or background field stars \citep{fei2,fei3}.  
We utilize this property to identify YSOs that do not have emission from a dusty disk
(evolutionary class III) and would otherwise be indistinguishable from field stars.  
Protostars (class 0/I) and pre-main sequence stars with disks (class II) with elevated 
X-ray emission may also be detected.
\citet{wol} observed the core of RCW 38 in a 96~ks observation.
The {\it Chandra} image of RCW 38 covered a 17$' \times$ 17$'$ field of view
centered on the cluster core and IRS2, thus X-ray data is not available over the entire 
spatial extent of the sources in our catalog.  
The 2006 catalogue contained 460 detections of which 294 were determined from 
their X-ray detection and NACO near-IR photometry \citep{der} to be likely young 
members of the cluster.  In this paper we make use of the reprocessed and updated 
catalogue, available on the ANCHORS online archive (AN archive of Chandra Observations 
of Regions of Star formation\footnote{http://cxc.harvard.edu/ANCHORS/}), to compare to the 
IR sources to identify likely cluster members.  

There were 536 X-ray detections in the catalogue, a mixture of YSOs, 
AGN, and dMe stars.  For each epoch, the coordinates of the X-ray sources 
were matched to the nearest IR detection to within a radius of 1$''$. 
In Epoch 1, 175 X-ray detections were identified with IR sources,  203 in Epoch 2, 190 
in Epoch 3, and 198 in Epoch 4.  In total,  226 X-ray sources were identified with IR sources.  
This number is less than the 294 X-ray detections matched in the \citet{wol} study since 
the resolution of the $Spitzer$ data is much lower than $Chandra$ in the central 5$'$ where 
$\sim$50\% of the X-ray sources are located.  
The lower limit of the X-ray luminosity detectable in RCW 38 with {\it Chandra} can be estimated 
by comparison with the COUP data, see \citet{fei}, as $log_{10}~L_X \approx 30.1$ $ergs$ $s^{-1}$, 
assuming a distance of 1.7~kpc to RCW 38,  an exposure time of 97~ks, and a median value of 
$log_{10}(N_H)$ of 22.3.     
We match the X-ray sources with IR counterparts to reduce the contamination from AGN; an 
X-ray source with no IR counterpart is assumed to be AGN contamination.  

Of those sources identified in X-rays, 97 were previously identified from the IR excess diagrams 
in Epoch 1, thus 78 YSOs were identified using X-rays.  In Epoch 2, 136 were identified in the IR 
and 67 were X-ray identifications. In Epoch 3, 110 were previously known, with 80 X-ray identifications.
In Epoch 4, 131 were identified as excess sources, with 67 detected in X-rays.  
In total, 74 YSOs were identified through their elevated X-ray emission.   
Of these, 41 had previously been identified in the \citet{wol} study, so that 33 are newly identified 
as members in this work.     
It should be noted that the ACIS field only partially covers the region, 
and thus the X-ray detection fractions are only representative of the central region and not 
the outer distributed YSO population.  Wider scale X-ray studies of region would allow us 
to examine the X-ray properties and dispersion of Class III objects in the region.    
The completeness of the X-ray data can be assessed from Fig.~\ref{complete}; 
it should be noted that only 67\% of the YSOs from class 0/I to class II lie in the $Chandra$ field 
of view, thus the completeness of the X-ray data is limited partially by the smaller field. 
Considering only those young stars located in the $Chandra$ field of view, and those YSOs with 
$m_{3.6} < 12$,  55\% of this sample are detected in X-rays.  
This percentage remains approximately constant over the range of  3.6$\mu$m magnitudes from 
$8 <  m_{3.6} < 12$.   Hence, we estimate the X-ray YSO detection completeness to be 55\% for 
objects with $m_{3.6} < 12.$.

\subsection{Summary of Identified Sources} 

The four epochs of data were combined, yielding a total of 679 candidate
members of the cluster over the IRAC overlap field.  Of these,  32 were found to be 
contaminants: AGN or PAH-rich galaxies, leaving 647 in the IRAC fields. 
The numbers of YSOs identified in each of epochs 1-4 were: 369, 358, 504, 377.   
On trimming this catalog to include only those sources in the IRAC overlap field 
of one or more epochs, the total number of cluster members was found to be 624. 
Of these, 226 were detected in X-rays by {\it Chandra}, including 74 that identified 
as YSOs by their elevated {\it Chandra} X-ray emission.  
The membership of RCW~38 is estimated to be 90\% complete to 12~mag at 
3.6~${\mu}m$, or $\sim$2~$M_{\sun}$ at 1~Myr \citep{sie}.  
As will be shown, $\sim$40\% of objects in each evolutionary class 
are detected in X-rays, thus we are possibly missing $\sim$200 class III members in the ACIS-field. 
We also identified a further 177 candidate cluster members, which would bring the total 
membership identified here to 801.  

By scaling the \citet{har} results to our 0.2~$deg^2$ field, we estimate that there is 
of order 1 AGB contaminant in our catalogue.  We estimate that 1 or less of the class III 
sources may be a dMe contaminant.

\section{\bf Evolutionary Classification}

\begin{figure}
\epsscale{1}
\plotone{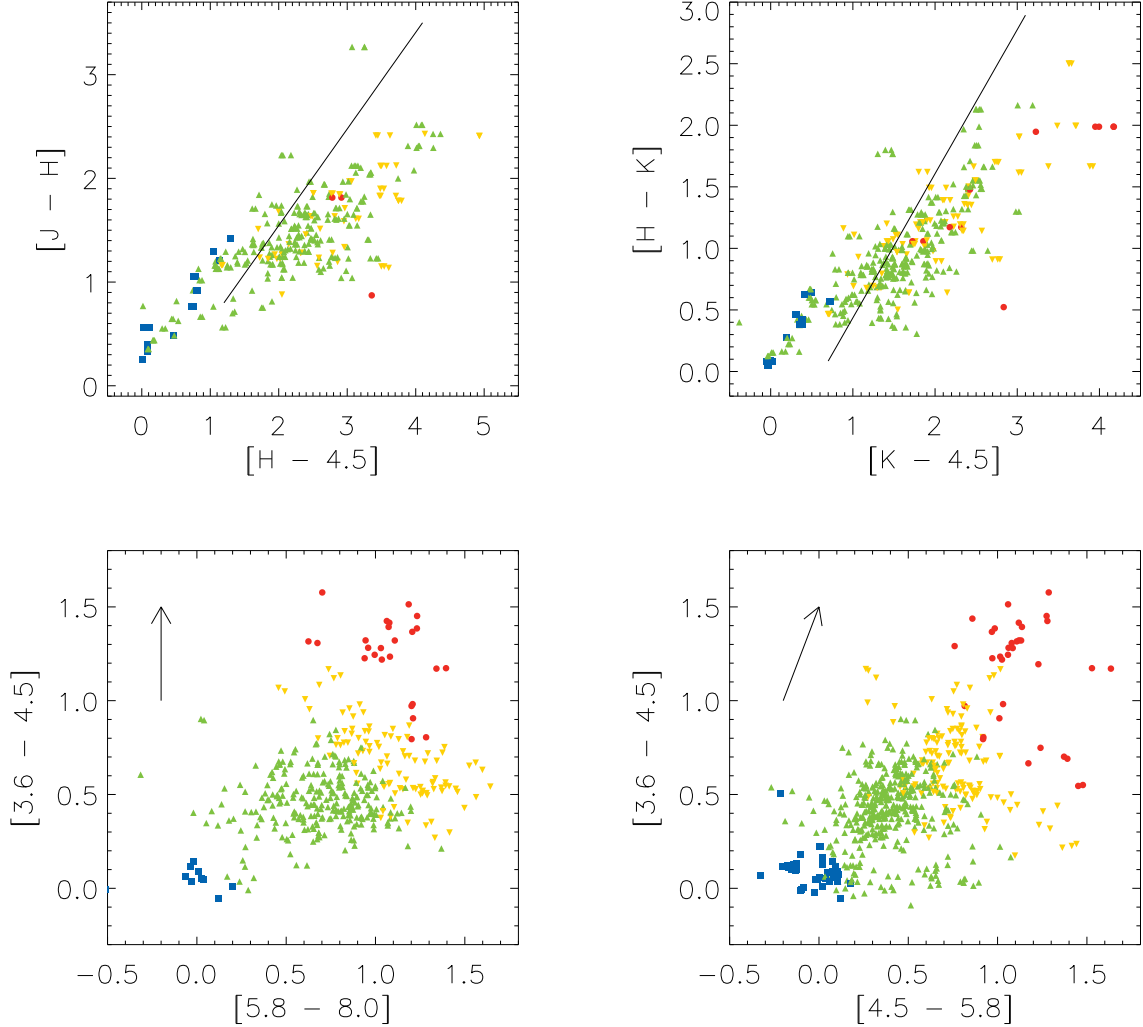}
\caption[]{ The same color-color diagrams as Fig.\ref{figccds}, in this case indicating 
the positions of the YSOs identified in our analysis.  The Class 0/I objects are marked 
by red circles, inverted yellow triangles mark the flat spectrum sources, green 
triangles mark the Class II objects.  The X-ray identified Class III are shown by blue 
squares.  On the IRAC color-color diagram some of the Class II stars appear to fall in the 
wrong region of the diagram for their class.  An examination of the dereddened SEDs of 
these sources shows that their positions on the color-color diagrams have been shifted 
by reddening.   }
\label{ccdsclass}
\end{figure}

The evolutionary classification of the sources was carried out by first noting their
locations on the IRAC and IRAC-2MASS color-color diagrams and 
assigning each source a preliminary class (\citet{winston,meg,all,gut09}, see Fig.~\ref{ccdsclass}). 
Typically, the IRAC color-color diagram was used for the initial assignment of the 
evolutionary class. 
This was followed by the construction of SEDs for all sources; where possible the 
dereddened SED was also constructed.   For each SED, a slope 
$\alpha_{irac} = dlog(\lambda F_{\lambda})/dlog(\lambda)$ was calculated.  
The conversion from magnitudes to fluxes in W~cm$^{-2}$~s$^{-1}$ used the 
following zero fluxes for the $J$, $H$, $Ks$, [3.6], [4.5], [5.8], and [8] bands
respectively: $3.13 \times 10^{-13}, 1.13 \times 10^{-13}, 4.28
\times 10^{-14}, 6.57 \times 10^{-15}, 2.65 \times 10^{-15}, 1.03
\times 10^{-15}, 3.02 \times 10^{-16}$ \citep{skr,faz}.
The slope was calculated by a least-squares fit over the available IRAC bands. 
The 2MASS bands are not included in the fit as they trace the stellar photosphere 
and tend to bias the fit towards shallower slopes, and thus an earlier evolutionary class.  
The initial classification from the CCDs is then refined by utilizing $\alpha_{irac}$, and 
$_{dered}\alpha_{irac}$ where available, to obtain a more precise classification.      

The merged master list of YSOs retained the classification assigned in each epoch 
to those sources detected in multiple epochs.  These classes were compared and 
found to agree in 485/624 cases (78\%).   In those were it did not, 139 cases (22\%),  
the class with the most IRAC detections was favored.  In 130/139 cases (94\%) the 
difference in class was between two neighboring evolutionary classes:  13 class 0/I / 
flat spectrum, 49 flat spectrum / class II, 68 class II / class III.  In the case of the class II 
and class III groups, longer wavelength observations would enable categorization of 
these objects as potential transition or anaemic disks.  Of the remaining nine objects, 
six were assigned evolutionary classes between class 0/I and class II, and the remaining 
three between flat spectrum / class II / class III. The latter three objects were checked 
and found to be lacking bands in particular epochs, resulting in spurious classifications. 

\begin{figure}
\epsscale{1}
\plotone{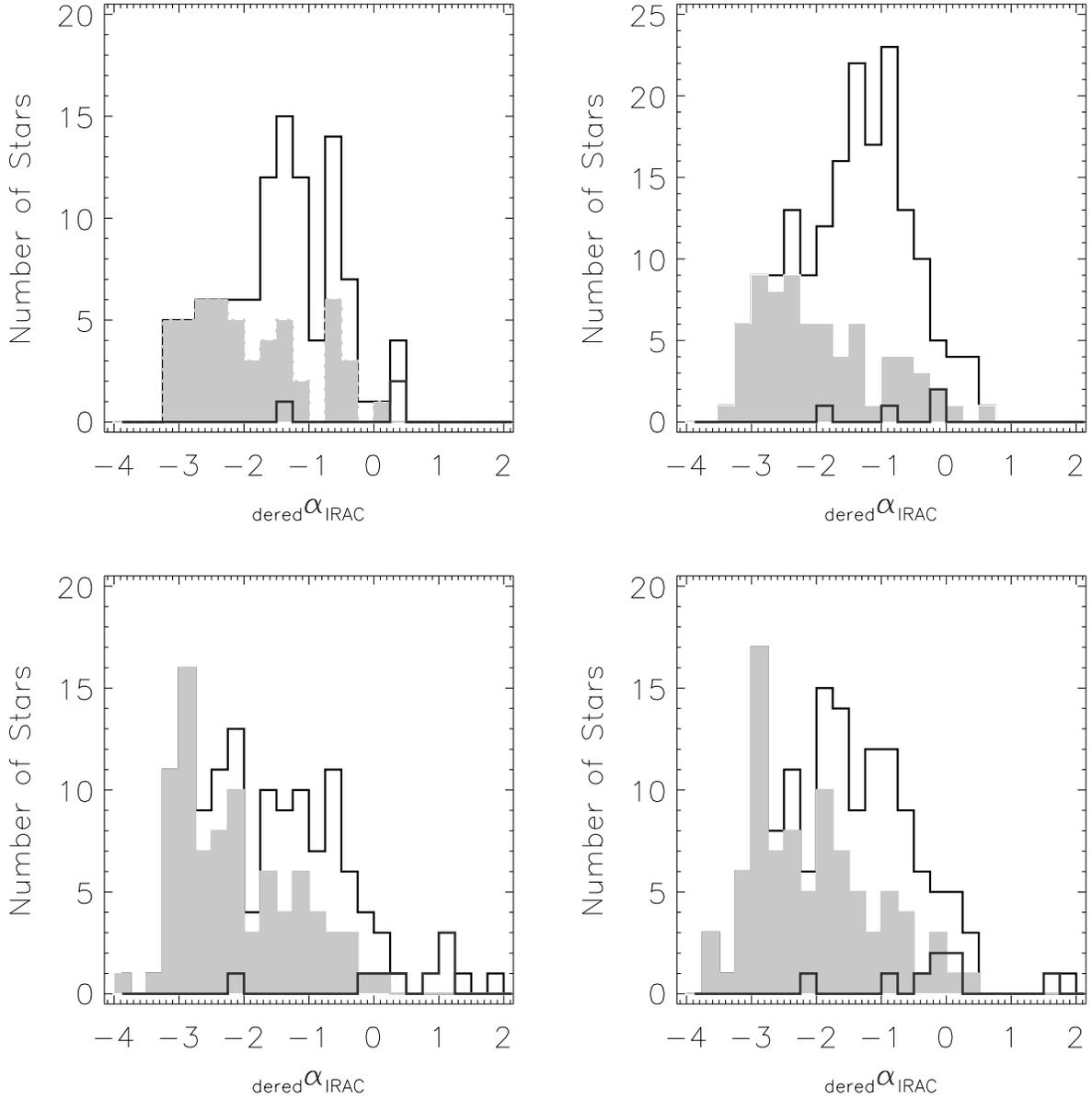}
\caption[]{Histograms of the dereddened SED slope, $_{dered}\alpha_{IRAC}$, of the 
identified cluster members of RCW~38.  
{\it Upper left:} The upper black histogram shows the identified YSOs in Epoch~1 with enough 
detections across the 2MASS and IRAC bandpasses to dereddened the photometry and to 
calculate $_{dered}\alpha_{IRAC}$. The gray filled histogram shows the subsample of the 
X-ray selected YSOs in Epoch 1.  The lower black histogram shows the candidate variable sources.  
{\it Lower left:} Similarly for Epoch 2.
{\it Upper right:} Similarly for the Epoch 3 YSOs.
{\it Lower right:} Similarly for Epoch 4.
}
\label{IRACslopes}
\end{figure}

In Fig.~\ref{IRACslopes} the dereddened IRAC slopes, $_{dered}\alpha_{irac}$, 
of the YSOs detected at each epoch are shown.  The open histograms 
represent the YSOs, the filled histograms show the 
X-ray selected sample at each epoch, and the lower black histograms 
indicate the variable sources at each epoch.   The median slopes of each 
epoch are similar; $-1.27\pm0.73$ (Epoch 1; $-1.21\pm0.73$, $-1.36\pm0.93$, $-1.29\pm0.88$, 
for Ep.2, 3, \& 4, respectively).  There is no indication of a difference in the median slope (and by extension 
evolutionary class) between the central core and the outer regions of the cluster.   
The X-ray sample have median slopes of $-2.19\pm0.90$ (Epoch 1; $-2.12\pm0.96$, $-2.33\pm0.94$, 
$-2.24\pm1.01$ for Ep. 2, 3, \& 4, respectively.), these are biased by the higher fraction of class IIIs included in this sample. 
The variable sources, which are discussed in Sec.~\ref{variability}, have median slopes of $-0.45\pm0.99$ 
(Epoch 1; $-0.59\pm1.05$, $-0.35\pm0.99$, $-0.29\pm0.99$ for Ep. 2, 3, \& 4, respectively).  The higher 
median value of the variables' slopes may be partially due to the smaller sample size, and is 
indicative of the younger nature of this YSO subsample.

A list of the coordinates and identifiers of the 624 sources identified as young stellar 
objects in RCW~38 are given in Table~\ref{tableastro1} to ~\ref{tableastro3}. 
Table~\ref{tableastro1} lists the coordinates, {\it Spitzer} ID, {\it Chandra} ID, evolutionary 
class, and variability indicator of the class 0/I protostars.  Table~\ref{tableastrof} list 
similar information for the flat spectrum sources, Table~\ref{tableastro2} the details of the 
class II objects, and Table~\ref{tableastro3} the information for the class III members.  
Similarly, Table~\ref{tablephot1} to ~\ref{tablephot3} lists the 2MASS and 
IRAC photometry for the identified YSOs. These tables  include 
measures of the apparent slope of the Spectral Energy Distributions (SEDs) and the extinction at 
$K_s$-band, $A_K$.

\section{\bf Discussion}\label{discussion}

There were 26,873 photometric sources detected in at least one IR band across 
all fields and epochs, of these 13,950 detections were located in the IRAC 4 band 
overlap region in at least one epoch. 
Using IR excess emission and X-ray detections, we have identified 624 young stellar objects in the 
IRAC four band overlap field of view. 
Of the 624 YSOs, 226 were detected in X-rays with {\it Chandra}.  
The YSOs were of the following evolutionary classes:  23 class 0/I, 90 flat spectrum, 
437 class II, and 74  class III.  Of the YSOs, 9 class I, 18 flat spectrum, 125 class II, 
and 74 class III were also identified in the {\it Chandra} X-ray observation.

In the following discussion we utilize this sample to address five specific topics. 
First, we compare the X-ray and IR properties of the YSOs detected with {\it Chandra}. 
Second, we compare the four epochs to determine if their properties vary and to identify variable members of the cluster.  
Third, we comment on the identification of 28 OB star candidates in the region.  
Fourth, deeply embedded objects identified in the MIPS observations are briefly discussed.   
Lastly, we examine the spatial distribution of the cluster YSOs as a function of their evolutionary class, variability and 
massive star content, and then discuss the implications for the underlying cluster structure.

\subsection{X-ray Characteristics}\label{xrc}

In recent years numerous studies have been undertaken to investigate the emission properties 
of pre-main sequence stars and protostars in the higher energy X-ray region of the spectrum
\citep{wol,get,win10}.   In developed, hydrogen burning stars, X-ray activity arises from 
magnetic fields generated as a result of shear between the core radiative zone and the outer 
convective zone.   In young stars, X-ray emission is thought to arise from mechanisms such as:  
magnetic disk-locking between the star and disk \citep{hay,iso,rom},  accretion onto the star 
\citep{kas,fav1,fav2}, or alternative dynamo models for coronal emission \citep{kuk,giam}.   
The elevated levels of X-ray emission in PMS stars has been shown to decrease 
with stellar age \citep{fei}, as is observed on the main sequence.

Of the 433 YSOs located within the Chandra field, 226 (52\%) had confirmed X-ray detections. 
Nine of these were class 0/I protostars (9/22 or 41\% of the class 0/I in the {\it Chandra} fov), 
eighteen were flat spectrum (18/58 or 31\% of the flat spectrum sources), 125 were class II 
detections (125/279 or 45\%). 
The remaining 74 sources did not exhibit excess emission and were identified as class III 
members of the cluster solely on their X-ray detection; hence 74/624 or 12\% of the members 
were identified solely by {\it Chandra}.  To within 1$\sigma$ there is no apparent trend in rate of
detection with evolutionary class, with a $\sim$40\% detection rate for all classes.  This detection 
rate is likely a function of the sensitivity of the {\it Chandra} observations.
Since we do not possess an independent method of identifying class III young stars we cannot 
establish the fraction of class III sources detected in X-rays.  As we shall subsequently argue 
the X-ray properties of class II and III objects appear to be similar, we therefore assume that the 
fraction of class III objects is also similar to that of the class II.  Here we neglect any suppression 
of X-ray activity in class II stars due to accretion \citep{tell}.  We thus estimate the total number of 
class III members to be $(74/44.8\pm4\%$ [percentage of CII detected in X-rays])$/60.5\%$ [correction 
for Chandra coverage] or 272$^{+26}_{-23}$, assuming a disk fraction of 67\% (see following discussion), 
for an estimated cluster membership of 823$^{+26}_{-23}$ objects with $mag_{3.6{\mu}m} > 12$.  

Of the 226 sources identified as young stars, 138 were previously identified as young members in 
the \citet{wol} study.  The remaining 88 consisted of 43 which were identified as YSOs by comparison 
to the IR catalogues in this study and 45 which were new X-ray detections included in this catalogue.   
Of the 74 class III sources, 41 were previously identified as members by \citet{wol}, the remaining 
33 class IIIs are newly identified in this study by their X-ray emission and IR photometry.  

The disk fraction of the YSOs,  the ratio of the number of X-ray detected disk-bearing pre-main 
sequence stars (class II) to the total number of X-ray detected pre-main sequence stars (class II \& III) was 
found to be $125/(125+74) = $ 63$\pm$6\%.  On including the X-ray detected protostellar sources (class 0/I \&  
flat spectrum) this fraction rises to $(9+18+125)/(9+18+125+74) = $ 67$\pm$5\%.   
The disk fractions are calculated based on the assumption of completeness to 12~mag at 3.6${\mu}m$, or 2~$M_{\sun}$, 
and similarity in the detection rates of class II and III sources.    
These fractions are consistent with a cluster age of $\sim$1~Myrs  \citep{her2}.  Both are significantly higher than 
the 29$\pm$3\% found by \citet{der} using VLT/NACO in a $\sim$0.5~pc area surrounding IRS~2.  
By considering only those of our sources falling in the same field (2 class II and 3 class III) we find a 
disk fraction, $2/2+3 =$  $40\pm28\%$.  Expanding to a 1~pc from the IRS~2 binary, the disk fraction 
becomes $17/17+7 =$ 71$\pm$17\%, which is comparable to the region as a whole.  These results suggest 
that proximity to the massive IRS~2 stars may have resulted an increased rate of disk processing in the 
surrounding lower mass YSOs.   

\begin{figure}
\epsscale{1}
\plotone{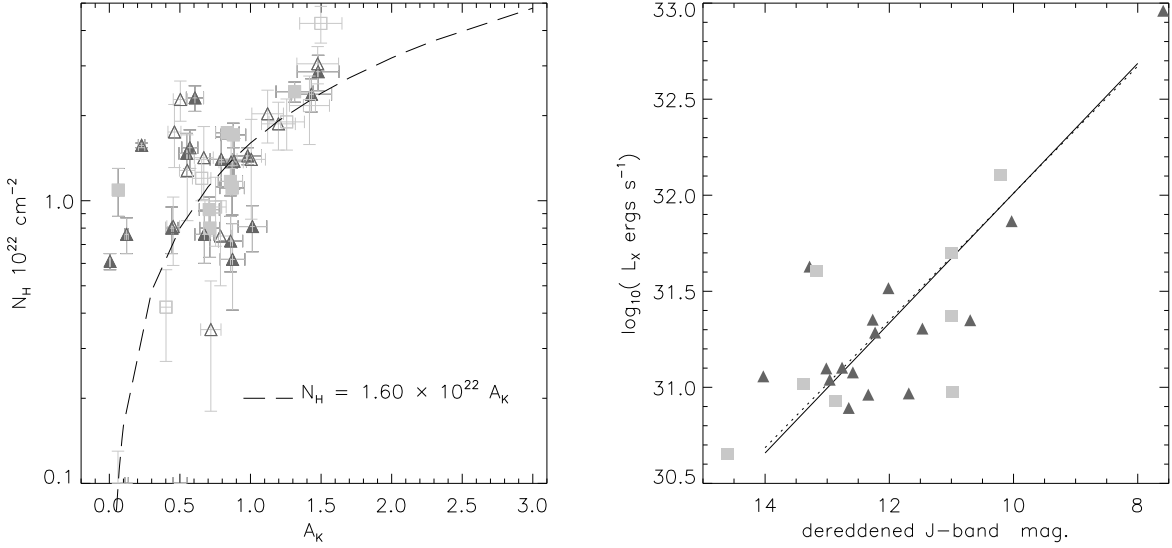}
\caption[]{ 
{\it Left:} The hydrogen column density, $N_H$, against extinction at $K$-band, $A_K$. The gas density was calculated 
from the {\it Chandra} data. The line indicates the standard $N_{H}/A_{K}$ ratio of 1.6$~\times~10^{22}$, 
showing a statistically good fit to the data.  
The filled symbols are class II (mid gray triangles), and class III (light gray squares) with $>$100 counts; the open 
symbols indicate sources with $>$50 counts. 
{\it Right:}  The X-ray luminosity, $L_X$, against the dereddened $J$-band magnitude (proxy for the bolometric 
luminosity).   No difference is observed between the X-ray luminosities of the class II (black triangles) and III (grey squares) 
sources with $J$ magnitude, the solid line indicates the linear fit to the class II objects, the dashed line the fit to the class III.  
 }
\label{xrays} 
\end{figure}

There are no discernible trends in $N_H$ against $kT$ with evolutionary class, the class II and III 
objects are similar. The class 0/I and flat spectrum sources have marginally higher column densities; 
this would be indicative of their surrounding infalling natal envelopes.  
The X-ray luminosity of pre-main sequence stars is known to vary with bolometric luminosity \citep{cas}.
As was previously found by the authors for Serpens and NGC 1333 \citep{winston,win10} and by 
\citet{her2} in $\sigma$~Ori, when using the $J$-band as a proxy for the bolometric luminosity in 
RCW 38 there is no difference in the dependence of $L_X$ on $L_{bol}$ between class II and III.  
Fig.~\ref{xrays}{\it(right)} shows the X-ray luminosity (at an assumed distance of 1.7~kpc) plotted against 
the dereddened $J$-band magnitude.    A linear fit to the 16 class II sources provides  
$log(L_X) = -0.338\pm0.027(J_{dered}) + 35.39\pm0.31$, while the fit to eight class II shows 
$log(L_X) = -0.331\pm0.042(J_{dered}) + 35.32\pm0.52$.  
This reinforces the suggestion that the mechanisms of X-ray generation are similar for both class II and III 
and is the basis for our estimate of the class III population. 
On the left of Fig.~\ref{xrays} the hydrogen column density is plotted against the extinction measured at 
$K_s$-band, $A_K$.   Only those detections with X-ray count in excess of 100 were considered when 
examining emission properties to ensure reliable estimates, though in Fig.~\ref{xrays}{\it(left)} the open 
symbols also show sources with $>$50 counts.

While this work was not explicitly designed to test gas absorption, $N_H$, versus dust extinction, 
$A_V  \approx  10 \times A_K$,  the X-ray data support the use of a simple model of an absorbed thermal 
plasma to measure $N_H$ to within 30\% (Wabs*APEC; \citet{mor,smith}).  
We can then examine the data for any trends in  the $N_H/A_K$ ratio.    
Historical measurements of $N_H/A_V$ range from approximately $2.2\times 10^{21} cm^{-2}$ \citep{ryt}  
derived from O-star absorption, to roughly $1.6\times 10^{21} cm^{-2}$ \citep{vuo} derived from a well 
behaved sample of PMS stars in the $\rho$ Ophichus cluster. 
Previously, in the lower mass clusters Serpens and NGC 1333, the authors reported on a 
decreased $N_H$ to $A_K$ ratio, of $\sim0.6 \times 10^{22}$ \citep{winston,win10}, than that 
quoted for the diffuse interstellar medium in \citet{vuo} of  $1.6 \times 10^{22}$ .   This is thought to be due 
to the agglomeration of grains in cold dense clouds.   However, in RCW~38 the sources are found to fit 
very well to the ratio of $1.6 \times 10^{22}$, consistent with the \citet{vuo} value for the local ISM and 
nearby molecular clouds, and with the previous measurement for the region of  $2.0 \times 10^{22}$ \citep{wol}.  
The class II have a median ratio of $N_H$$= 1.66\pm0.85\times10^{22}$$A_K$ from sixteen sources, 
while the class III have a median ratio of $N_H$$= 1.52\pm0.49\times10^{22}$$A_K$ from eight objects.       
The difference may arise from the much higher mass of the RCW~38 cluster.  \citet{smi} found that the gas to dust 
mass ratio towards IRS~1 was 10$^4$:1, indicating dust depletion with respect to the ISM.  
The more active stars may efficiently irradiate the dust in the intercluster medium 
preventing grain growth or aggregation of dust grains, which would be more likely in the colder 
medium of the less massive clusters \citep{card,step}.  A larger scale survey of the region, including the outer 
subclusters, would allow us to ascertain if this difference is more dependent on nascent environment 
or on the presence of OB stars in the vicinity.

\subsection{Variability}\label{variability}

Young stars vary photometrically over time.  One of the first criteria, in fact, by which young 
T Tauri stars were identified was their photometric variability.  This variability occurs across 
all wavelengths from the UV through optical to IR and into the high energy X-ray regime.  
The causes are manifold; XU Ori / FU Ori flaring, stellar spots, stellar pulsation, and in the 
IR, heating of the circumstellar disk  \citep{carp}.  
In RCW~38 we have four epochs of observations to search for known YSOs exhibiting 
variability and for young diskless class III members not detected in X-rays.  
Given the different regions and exposure times of the outer field and central  
pointings and the consecutive observations of epochs 1 \& 3 and 2 \& 4, only the 
variance of sources between Epochs 1 \& 2 and 3 \& 4 were examined.  Further, only the 
IRAC 4 band overlap subsample was considered.  Thus we are searching for variability on 
time scales of $\sim$1 year.     
A variable source was defined as one having a difference in magnitude in one or more 
IRAC bands of $>3\sigma$ above the combined uncertainties.   

\begin{figure}
\epsscale{1}
\plotone{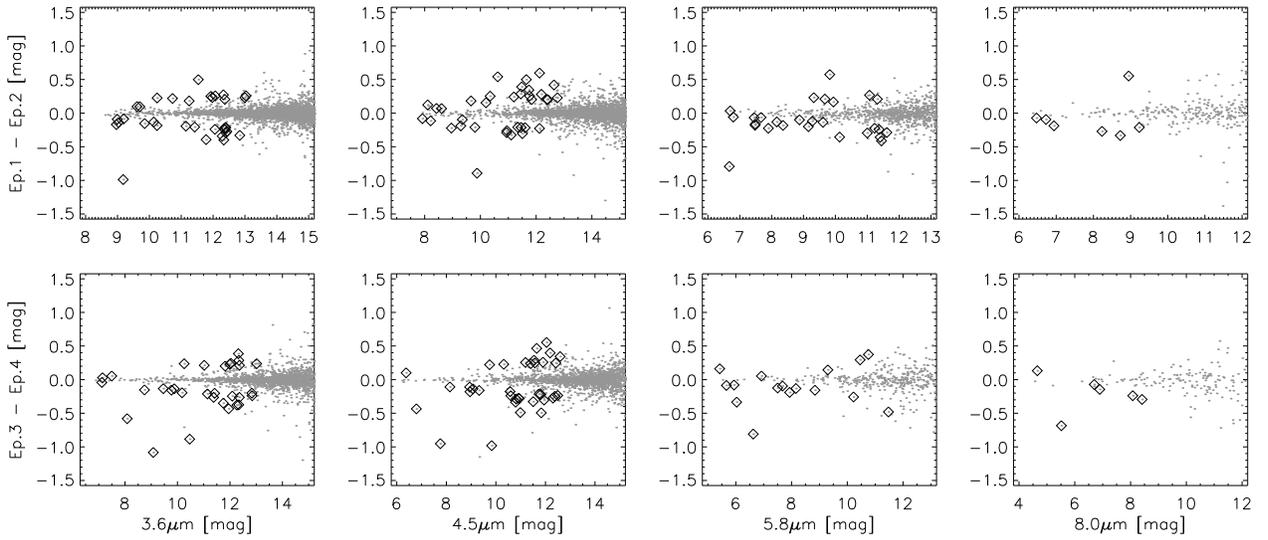}
\caption[]{
Variance with magnitude of the 72 candidate variables in RCW 38 over the four IRAC bandpasses. 
In the identification of variable candidates, only the comparison of photometry from [Ep.1 - Ep.2] and 
[Ep.3 - Ep.4]  was considered as Epochs 1 \& 2 are the larger fields surrounding the core and have longer 
integration times, and Epochs 3 \& 4 are the exposures of RCW 38 core. 
Sources were selected as variables if they exhibited a magnitude difference in any band greater than 3$\sigma$.  
Only sources with uncertainties $<0.1$~mag were considered.     }
\label{varymags}
\end{figure}

Of the 13950 sources in our sample, we identified 72 candidate variable stars.  
Fig.~\ref{varymags} shows the variation of the 72 variable stars at each of the four 
IRAC bandpasses for the two timeframes considered. 
Forty-eight of these were previously identified as YSOs due to excess emission; variability 
is noted in the final columns of Tables~\ref{tableastro1}-\ref{tableastro2}.  
The remaining 24 objects were not otherwise found to be YSOs; from an examination of their 
SEDs it was found that two exhibit excess emission characteristic of circumstellar disks but 
were not detected in enough bands to be placed on the color-color diagrams, 
while the other 22 are candidates to be diskless class III young stars.   The coordinates, identifiers 
and photometry of these twenty-four sources are given in Table~\ref{tablevariables}.  
Of the 24 objects, nine were found to also be O-star candidates and are marked as such in 
Table~\ref{tablevariables}.  
Of those 48 previously identified YSOs found to be variable, seven were class 0/I (three X-ray sources), twelve flat 
spectrum (4 X-ray), and twenty-nine were class II (7 X-ray).  No known class III were identified 
as variable.  These numbers represent fractions of variables per class of 30$\pm$11\% 
 for class I, 14$\pm$4\%  for flat spectrum, and 7$\pm$1\%  for class II.  The variability fractions for 
 the X-ray detected sample of YSOs was consistent with that of the non-X-ray detected sample. 
The protostellar objects have a combined variability fraction of 18$\pm$4\%   
compared to the 7$\pm$1\%  for the disk-bearing class IIs.  Such high variability 
is consistent with larger and/or more frequent accretion events during the protostellar phase.

A study of the properties of these variable sources found that they were evenly spatially 
distributed across the IRAC overlap field; this is not unexpected as variability is a universal 
property of young stars.  The variables exhibit a similar range of extinctions as the YSOs not 
found to be varying and there was no trend found between the extinction at $K$-band and the 
extent of the variance.  There was no statistically significant difference in detection of variance in 
X-ray luminous objects or in the X-ray luminosity of the variable members.  
There is some indication that the percentage of variables decreases with evolutionary class, with 
$\sim$20\% of the protostellar YSOs varying compared with $\sim$7\% of the class II. None of the 
previously detected class III sources were found to vary.  
The extent of variance also appears to show some decrease with evolutionary class:  the mean 
variance of the class 0/I was 0.43$\pm$0.31, that of the flat spectrum 0.26$\pm$0.16, the class II 
0.27$\pm$0.17, and those variables showing no infrared excess emission 0.12$\pm$0.12. The 
trend is not strongly significant but is consistent with a decrease in the scale of accretion events 
in the later evolutionary stages.

\subsection{Candidate O Stars}

RCW~38 is one of the nearest massive star forming regions with an estimated membership 
of more than 1000 members.  Previously, \citet{wol} identified 31 candidate OB stars in a 5~pc$^2$ 
region centered on (and including) IRS2 using an absolute $K$-band magnitude $\ge$-0.35, for a 
2.7~$M_{\sun}$ at 0.5~Myr \citep{sie}. Twenty were located in the central $\sim$1~pc.  
Following \citet{knod} we use the $K$ v. $J - K$ color-magnitude diagram to identify the massive 
stars in the subsample of objects within the IRAC 4-band overlap region.   
We identify 29 candidate O stars and a total of 604 OB star candidates.   The coordinates and photometry 
of the 29 O star candidate sources are given in Table~\ref{tableOstars}.  
Of the 29 objects, nine were found to overlap with the variable sources and are marked as such in 
Table~\ref{tableOstars}.      An off-field region was subtracted from the cluster to account for field star 
contamination in the cluster; subtracting the off-field sources yielded an estimate of 26 
O stars in the RCW 38 region; thus we estimate that $\sim$~3 of our 29 objects may be contaminants.  
Two of the objects, 158 and 244, are identified as YSOs and are assigned the evolutionary class II, they are 
X-ray detected and were also previously identified as candidates by \citet{wol} as 112 and 396, 
respectively.  The small overlap (2 of 31) between the two lists arises from two causes.   Firstly, the smaller 
field of view of the \citet{wol} VLT/NACO observations and their higher sensitivity in the cluster centre, 
where we require a detection in at least one of the IRAC bandpasses.   Secondly,  the \citet{wol} study 
searches for OB star candidates with $M_{K} \le -0.35$, while here $M_{K} \le -3.65$.   
Of the remaining 31 sources, a further nineteen are classified as YSOs, and two more had IRAC detections in 
too few bands to classify.  Seven were not detected as point sources with IRAC.  IRS~2, the known O5.5 binary, 
is not resolved in the IRAC mosaics and does not appear in the candidate list.         

Previously, the total population of RCW~38 was estimated as being of order one thousand members.  
An estimate of the total membership of the RCW~38 region has been presented in Sec.~\ref{xrc} as   
823$^{+26}_{-23}$ objects with $mag_{3.6{\mu}m} > 12$.   In comparison,  \citet{wol11} estimate the total 
population of Trumpler~16, which contains a similar number of massive stars,  as being roughly 
14,000 young stars.   From this we suggest that the total population of young stars forming in the 
RCW~38 region is closer to $10^4$, making it one of the more massive known clusters within 2kpc.

\subsection{Deeply Embedded Objects \label{deo}}

The MIPS 24~$\mu$m mosaic was examined in a search for embedded protostars 
or young objects associated with emission regions.  Three such emission regions 
were identified to the N of the cluster along the ridge of pre-main sequence stars 
forming the NW subcluster.  The diffuse emission is associated with three stars 
not previously identified as YSOs, variable or O-star candidates.  Their IRAC colors 
and SEDs suggest that they lack circumstellar material and that the emission arises 
from dust heating of the surrounding nebula.  The coordinates and photometry of 
the three stars are given in Table~\ref{table24}.    

A deeply embedded object was detected in the central subcluster, at coordinates 
$8^{hr}59^{m}05.95^{s}$, $-47^{d}30^{m}39.7^{s}$.  It is offset from IRS~2 by 
$\sim$3.5" or $\sim$6000~AU.   This object was detected only in the 8.0~$\mu$m band 
(no 24${\mu}m$ photometry is available due to saturation); it does not alter its position 
during the year between the Ep.2 and Ep.4 observations and hence it is assumed to be 
a deeply embedded protostar.  It has a magnitude of 4.679$\pm$0.099 at 8.0~$\mu$m 
and is faintly visible at 5.8~$\mu$m, though not resolved as a point source.  It is not 
detected at 4.5~$\mu$m suggesting an absence of shocked molecular hydrogen 
emission, which would indicate that it is not a jet feature.   Therefore we assume that 
this object is an embedded early protostar and may be embedded behind the O5.5 binary.   
A second object was also identified only at 8.0~$\mu$m but was found to move 
from Ep.2 to Ep.4 and hence considered a contaminating foreground source.

\subsection{Subclustering and YSO Spatial Distribution}

The spatial distribution of young stellar objects in a cluster provides an 
insight into the fragmentation processes leading to the formation of
protostellar cores, evidence for triggered star formation and the subsequent 
dynamical evolution of the stars as they evolve from the protostellar to the 
main sequence \citep{all2}.  
While \citet{der} examined the young stars in the inner core of the cluster 
at near-IR wavelengths, finding no evidence of subclustering, we will examine 
the extended molecular cloud. By including the longer mid-IR wavelengths we 
can obtain a clearer determination of the IR excess emission, and thereby more 
accurately classify the YSOs as protostellar, disk-bearing class II or 
diskless class III.  These data provide a deeper understanding of the star 
formation history of the extended RCW~38 region.

\subsubsection{Spatial Distribution of Identified YSOs}

The spatial distribution of the young stars in the cluster in each evolutionary 
class was examined using a nearest neighbor technique.  We define the nearest neighbor 
distance as the projected distance to the nearest YSO of the {\it same} class, assuming 
a distance to RCW~38 of 1.7~kpc.  
For each evolutionary class, 10K random distributions of stars were generated, with 
the number of stars equal to the number of objects in the given evolutionary class.  
For the class III sources, the randomly distributed stars were constrained to fall within 
the {\it Chandra} field of view, for all other classes, the random distribution covered the region 
of the IR-field. The resulting nearest neighbor distributions of the observed YSOs and 
the random distributions were compared for each class using the K-S test.  The 
probabilities that the random and observed distributions were drawn from the same 
parent distribution are listed in Table~\ref{tableprobs}.  The distributions of the class 0/I, 
flat spectrum and class II are found to be non-random.  The class III sources show some 
evidence of mixing; this is likely a statistical effect of the smaller $Chandra$ field of view.

\begin{figure}
\epsscale{0.75}
\plotone{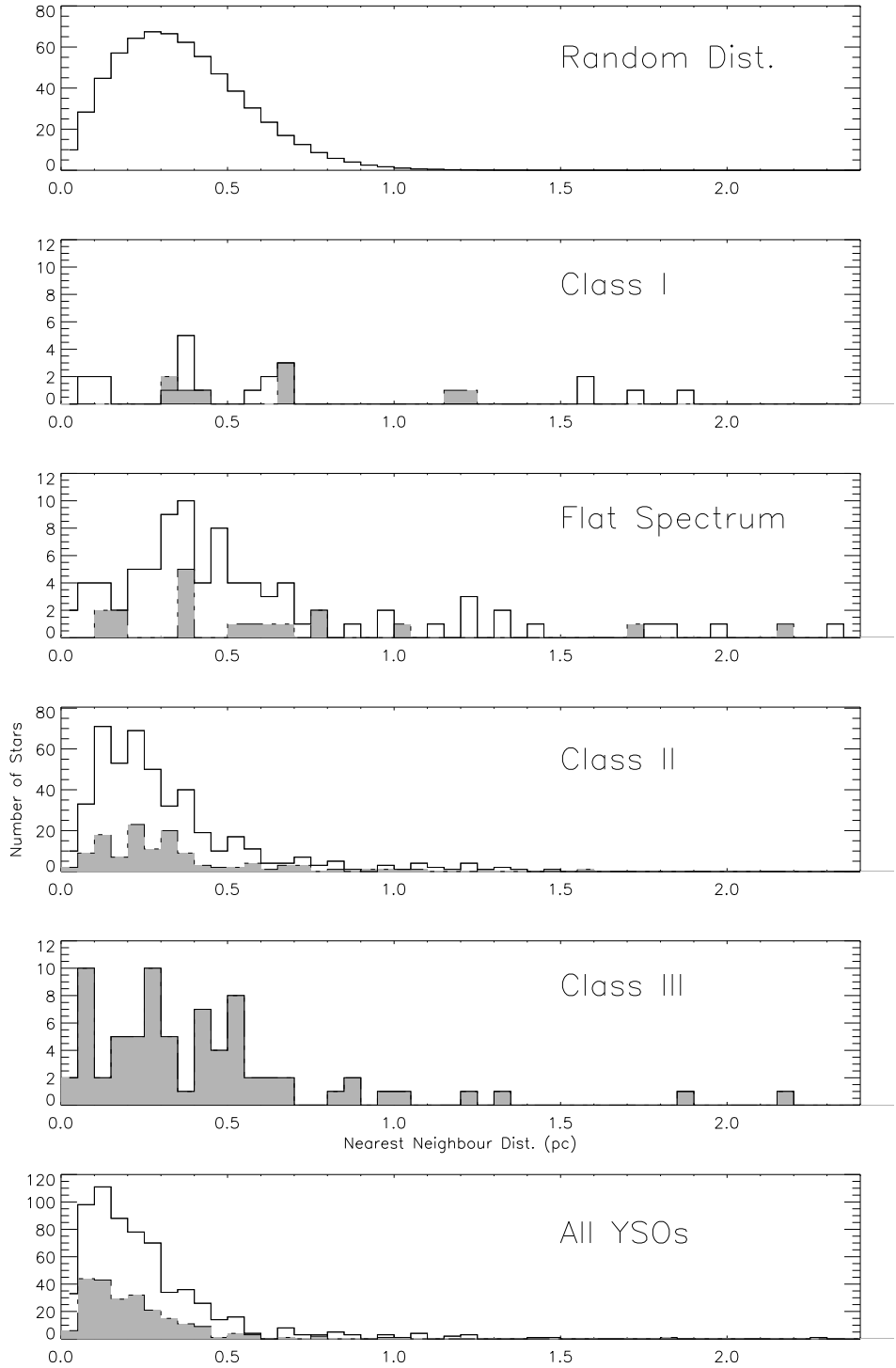}
\caption[]{
Differential distribution of nearest neighbor distances by class for all the identified YSOs, 
shown by the solid black line; the shaded region gives the differential distribution of nearest 
neighbor distances by class for the X-ray selected sample of all YSOs. The stellar sources are 
more densely clustered than the protostellar objects.  
The top panel gives the nearest neighbor distances for a random distribution.   
}‡
\label{nnplots}
\end{figure}

Fig.~\ref{nnplots} shows the nearest neighbor differential distributions by evolutionary 
class of the young stars.  The open histograms show all sources in that class, while the 
filled histograms show the subsample of X-ray luminous stars.  
The upper plot shows the  random distribution, averaged over the 10K runs.  
The median spacing of the protostellar members was larger than that of the pre-main 
sequence objects.  The median separation over the field for the class 0/I sources was 0.644~pc, 
and of the flat spectrum 0.433~pc.  In comparison the pre-main sequence class II members 
had a median spacing of 0.249~pc, and the class III 0.346~pc.

\begin{figure}
\epsscale{1.}
\plotone{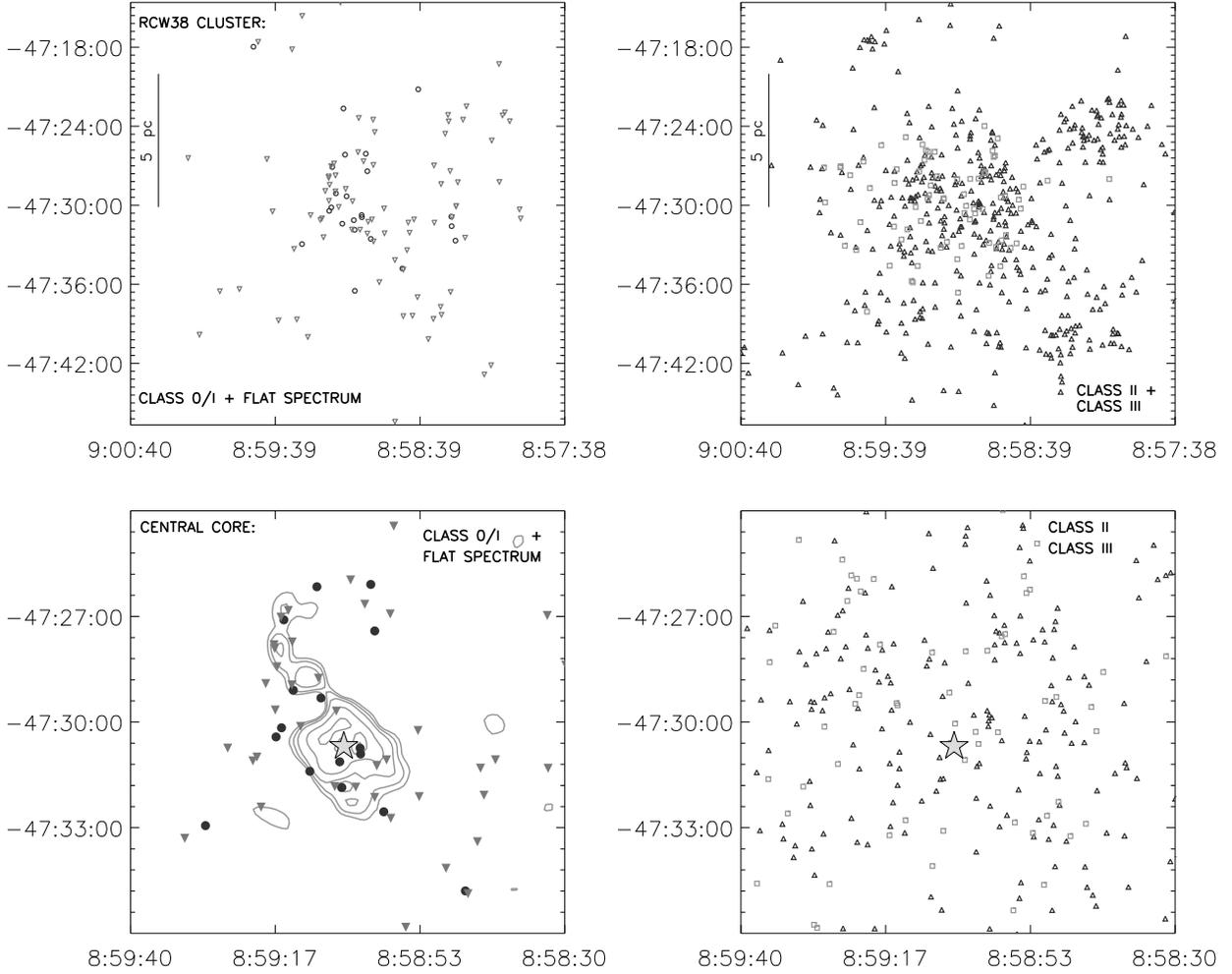}
\caption[]{
Spatial distribution of the various classes of YSOs in the RCW 38 cluster.  
{\it Above Left:} Class 0/I  and flat spectrum sources, shown by circles and inverted triangles, respectively.  
{\it Above Right:} Class II objects and class III sources, shown by triangles and squares, respectively.  
Evidence of subclustering can be seen from the distribution of CII/III sources and the distribution 
of protostellar objects.  Note that the class III distribution is delimited by the {\it Chandra} field of view, which 
is smaller than that of IRAC.  
Expanded view of the YSOs in the centre of the RCW 38 cluster, in a $\sim$0.2 deg diameter of the central O star binary.  
{\it Below left:} Class 0/I  and flat spectrum sources, shown by circles and inverted triangles, respectively.  
The \citet{vig} SIMBA 1.2mm map of the region is shown in contours, with contour levels of [58\%, 59\%, 61\%, 70\%, 85\%, 95\%].  
{\it Below right:} Class II objects, shown by triangles, and class III sources, shown by squares.  
The stellar members appear to be evenly distributed, while the protostellar objects form an arc roughly 
centered on the O star binary, IRS~2. The protostars appear coincident with the denser dust.     
}
\label{sdf}
\end{figure}

Fig.~\ref{sdf} shows the spatial distribution of all identified YSOs as a function of 
their evolutionary class in the upper plots.  The lower plots shows the spatial 
distribution of the young stars, again as a function of evolutionary class, in an expanded 
region centered on IRS~2.   The detection of class III members is limited 
by the smaller field of view of the {\it Chandra} ACIS instrument.   
Evidence of subclustering and separate protostellar formation sites  can be observed in 
the upper panels of Fig.~\ref{sdf} - indicating that a unique centre of star formation is unlikely. 
An arc or filament of protostellar objects is also observed surrounding the central IRS~2 cavity.  
The formation of these young stars may have been 
influenced by their proximity to the IRS~2 binary.  The spatial distributions of the class II and 
III stars surrounding IRS~2 do not trace this arc/filamentary structure.  A SIMBA 1.2mm map 
of the IRS~2 region is overplotted in the lower left of Fig.~\ref{sdf}, and shows a ridge of emission 
which is spatially coincident with the filament of protostars \citep{vig}.   This implies that the protostars 
remain embedded in their natal, dense dust and gas while the pre-main sequence stars have 
dispersed through the cluster.

\subsubsection{Subclustering: Surface Density}

\begin{figure}
\epsscale{1}
\plotone{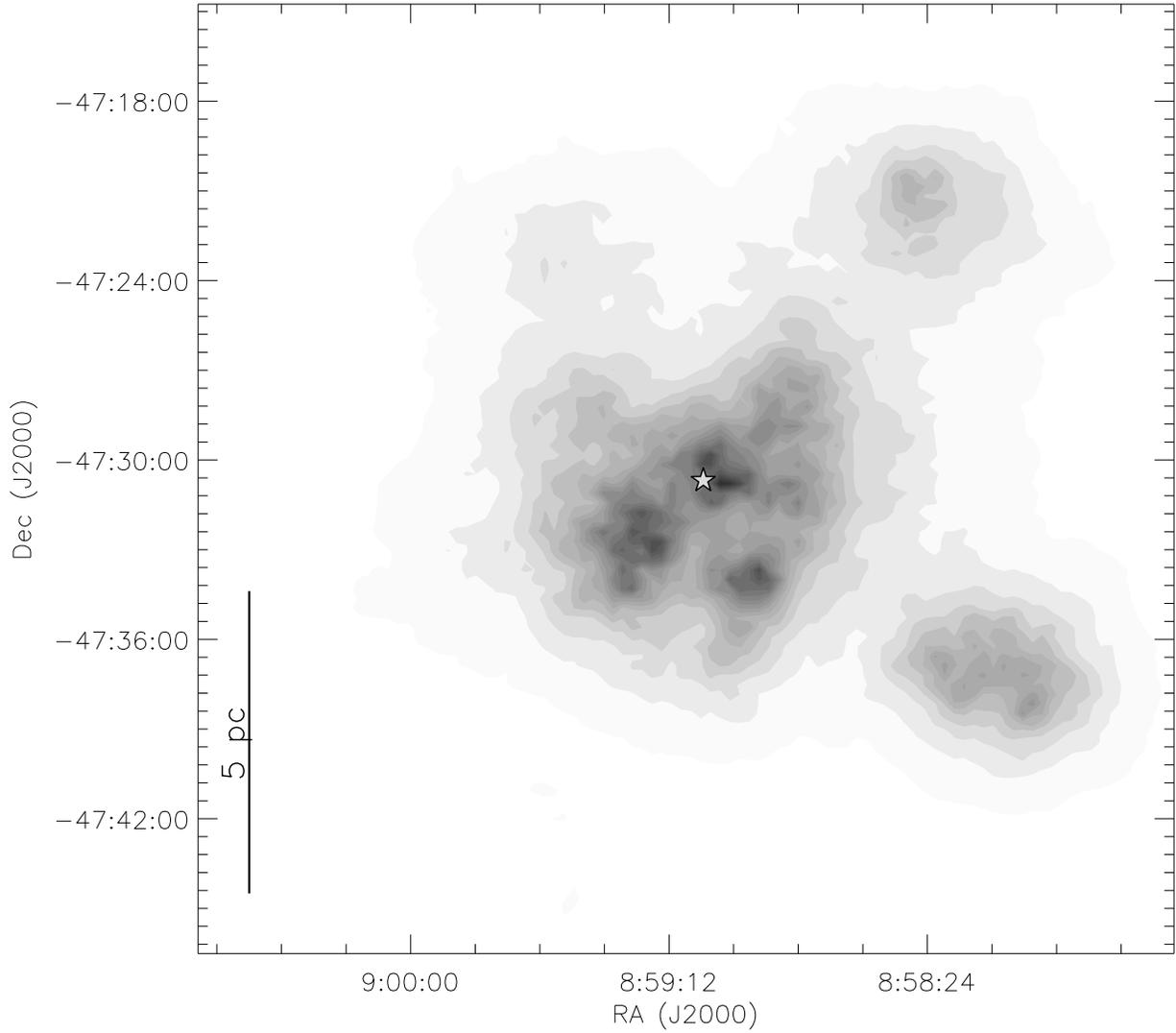}
\caption[]{
Surface density map of the YSOs in the RCW~38 cluster.  The stellar surface density was calculated with a grid size 
of $100\times100$ using $N = 18$, the distance to the 18th nearest neighbor, to smooth out smaller scale structure.   
The density contours range linearly from $0 / pc^2$ (white) to $30 / pc^2$ (black).  The surface density map also 
indicates the presence of two subclusters to the NW and SW, and a possible third subcluster to the NE, of the central 
core, which itself shows three density peaks.    The position of IRS~2 is marked by a star. 
}
\label{ysodensity}
\end{figure}

The stellar surface density of the YSOs was also calculated to determine the underlying 
structure of the cluster; the resulting map is shown in Fig.~\ref{ysodensity}.  
The local surface density at each point on a uniform grid was calculated following the method 
outlined in \citet{case}:
\begin{eqnarray*}
\sigma(i,j) = \frac{N-1}{\pi~r_{N}^{2}(i,j)}
\end{eqnarray*}
\noindent where $r_N$ is the projected distance to the $Nth$ nearest cluster member.  
In this case, the stellar surface density was calculated with a grid size of $100\times100$ using 
$N = 18$, the distance to the 18th nearest neighbor, to smooth out smaller scale structure.  
Further, \citet{case} found that the uncertainty goes as $\sigma/(N-2)^{1/2}$, so by taking $N=18$, 
the uncertainty is 25\%.    The surface density plots show three subclusters surrounding the main 
IRS~2 core:  a large subcluster to the SW, one to the NW, and a possible small grouping to the NE.  
The central cluster itself appears to have three density peaks which are likely real, though small, 
overdensities in the YSO population.  One of these peaks corresponds to the region immediately 
surrounding IRS~2.    The central subcluster coincides with the eastern cloud of \citet{yama}.  
The SW subcluster is coincident with the western cloud  implying continued star formation in that region.

\subsubsection{Subclustering: Minimum Spanning Tree}

\begin{figure}
\epsscale{0.75}
\plotone{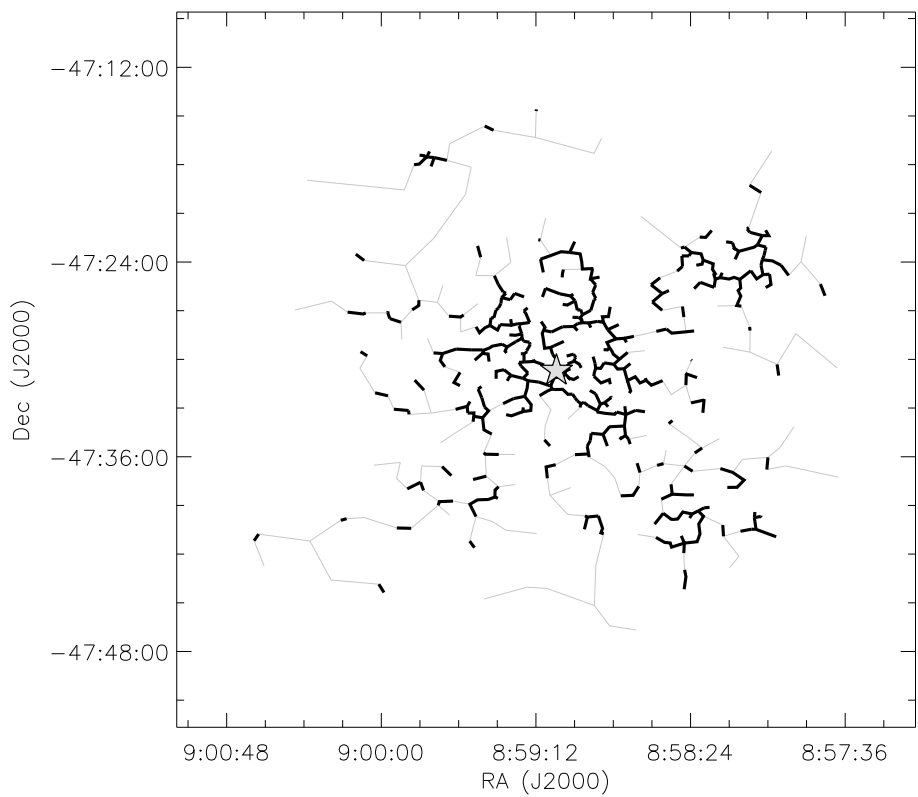}
\caption[]{ Minimum spanning tree (MST) of the YSOs detected in the RCW 38 cluster. 
The vertices represent the positions of the YSOs.  The black branches are those with a length less 
than the characteristic branch length of 50$"$, c.f. Fig.~\ref{mstbl}.  The grey branches have lengths 
greater than 50$"$.  Those YSOs connected by branch lengths shorter than the characteristic length 
can be considered to form a cluster.  The MST indicates that the RCW~38 cluster can be 
subdivided into the central core, focused on the O-star binary, and two separate subclusters to the 
NW and SW. There is also a small grouping of eight stars to the NE, which may be a developing 
subcluster.   The position of IRS~2 is marked by a star.   
}
\label{mst} 
\end{figure}

\begin{figure}
\epsscale{0.75}
\plotone{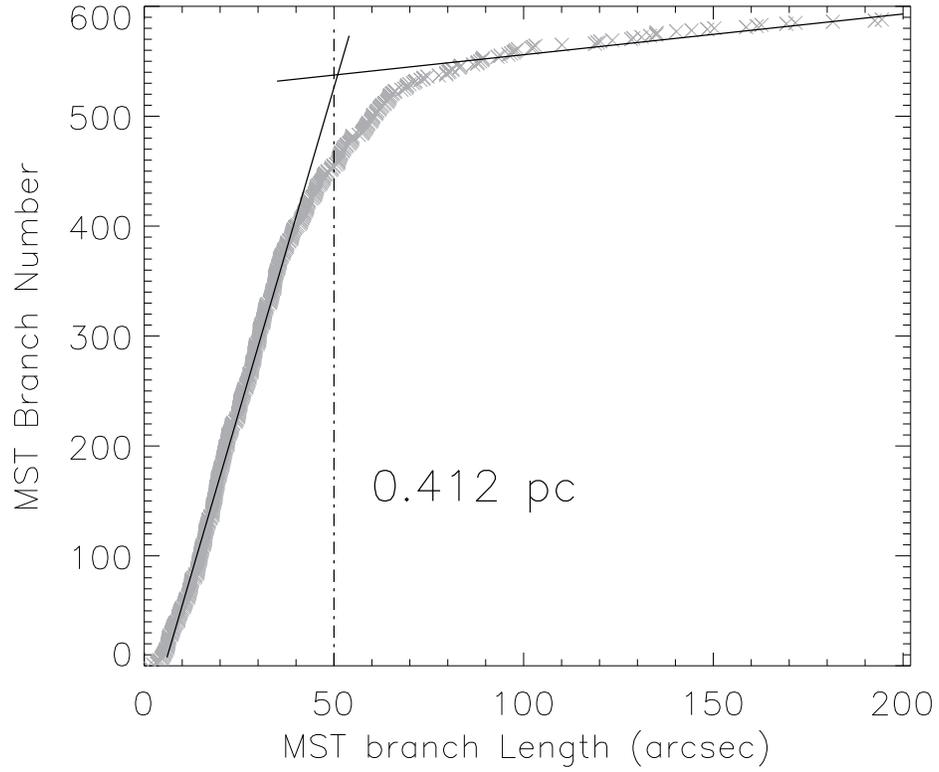}
\caption[]{
Distribution of branch lengths for the minimum spanning tree of the RCW~38 YSOs. The distribution
was fitted in two linear regimes. The characteristic branch length for the cluster 
was found to be 50$"$ or 0.412~pc.
}
\label{mstbl}
\end{figure}

Another method of examining the spatial distribution and structure of a cluster is the 
minimum spanning tree (MST). The MST is defined as the set of branches or lines 
connecting a set of points such that the sum of the branch lengths is minimized and 
there are no closed loops.  Fig.~\ref{mst} shows the MST for the YSOs in RCW~38. 
The MST provides a better method of characterizing the separation lengths of sources 
as all points are connected in a single network, whereas the nearest neighbor technique 
tends to create small subgroupings due to pairing of adjacent points. The MST also 
has the advantage of not smoothing out geometry in the substructures.  By assuming a 
critical MST branch length, the length where a break occurs in the cumulative distribution 
of branch lengths, substructures within the cluster can be defined.
Following \citet{gut09}, in Fig.~\ref{mstbl} we plot the distribution of branch lengths in 
the MST to obtain the characteristic branch length, found to be 0.412~pc. This was done 
by applying two linear fits to the data using the IDL routine $linfit$ and minimizing the $\chi^2$ 
uncertainty of the two fits.  In Fig.~\ref{mst} the black 
branches are those with branch lengths less than the characteristic branch lengths 
while the gray branches are those with longer lengths.  Those stars joined by black 
branches can be said to trace subclusters with the extended region.  We note that 
the central core has two large subclusters to the NW and SW, and a third smaller 
subcluster to the NE, containing only eight stars, which may be a developing region.

Extending the usage of the MST, \citet{car} formulated the $Q$-parameter, to quantify and 
distinguish between smooth large-scale radial density gradients and multiscale or 
fractal subclustering.  This method provides a statistical method of characterizing structure 
in stellar clusters.  The $Q$-parameter is defined as $\bar{m} / \bar{s}$, 
where $\bar{m}$ is the normalized mean MST branch length, 
$\bar{m}(N_{total}A)^{1/2}/(N_{total}-1)$, and $\bar{s}$ is the normalized mean 
separation,  $\bar{s} / R_{cluster}$. A $Q$-parameter equal to $1$ indicates a smooth 
radial distribution of sources, while values closer to zero indicate subclustering.  
For the RCW~38 region observed with IRAC, $\bar{m} =$ 0.2219~pc and 
$\bar{s} = $ 0.5407~pc, yielding a value of $Q$ equal to 0.41.  This low value 
indicates that subclustering defines the structure and spatial distribution 
of the YSOs.   \citet{der} found a value of $Q$ in the smaller central IRS~2 region of 0.84, indicating 
that no subclustering was present on this scale.  This result is consistent with our own since the \cite{der} 
study covered an area of  $\sim0.5\times0.5~pc^2$, corresponding to the density peak containing IRS~2 
within the central subcluster.  On this smaller scale, any structure may be masked by the dynamical 
evolution of the young stars. On the larger scale the underlying subclustering of the complex is evident.

\subsubsection{The Subclusters}

The RCW~38 region has extended star formation over the 30'$\times$30' IRAC field, a larger area 
than previously assumed.  The central cluster where IRS~1 \& 2 are located is surrounded by three 
subclusters and distributed star formation.   Table~\ref{tablesubcs} presents the populations of the 
subclusters,   their median extinction and IRAC slope, and their protostellar and variable fractions.

\begin{figure}
\epsscale{1}
\plotone{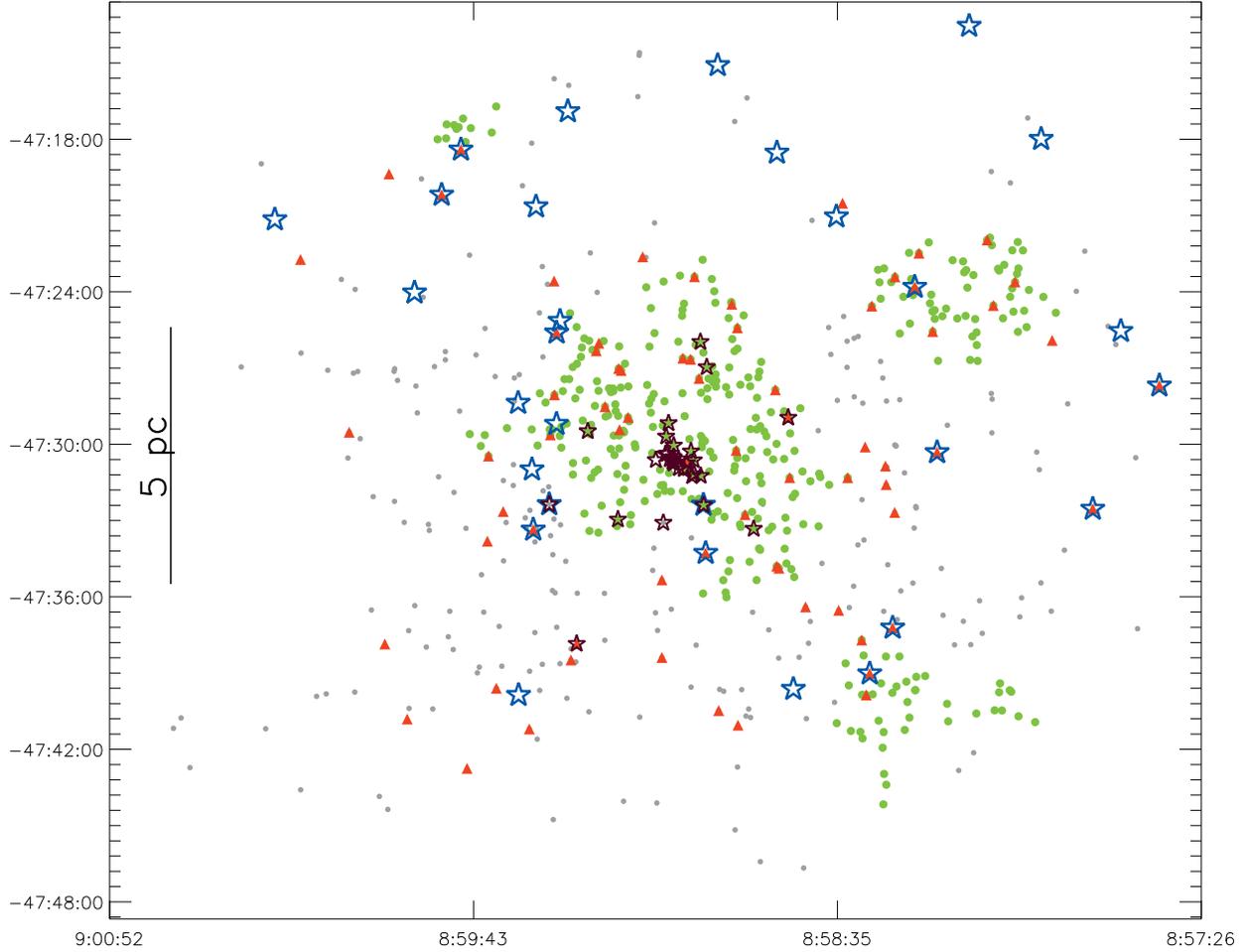}
\caption[]{
Spatial distribution of the identified YSOs in RCW~38 showing the positions of the O star 
and variable candidates.  The dots indicate the young stars identified via IR excess 
emission and X-ray detection; clustered YSOs are shown by green dots, distributed YSOs in gray.  
The triangles show the positions of the variable objects, while the blue open stars show the positions of 
the candidate O stars in the field.   The black open stars show the positions of the candidate OB stars 
identified in \citet{wol}.     Both the variable and O star candidates are distributed throughout 
the extended cluster, with candidates of both types appearing in the subclusters.    
}
\label{obvar}
\end{figure}

Each of the subclusters contains variable stars; the NE subcluster contains just one, the NW 
contains eight, the SW three, and the central subcluster contains twenty-four.  The remaining 36 
form part of the distributed population.  While the four subclusters contain very similar fractions 
of variables, the distributed population contains $\sim$50\% more.  
The protostellar fractions of the NE and NW subclusters and the distributed population are similar at 
$\sim$18\%, while the the SW subcluster fraction is appreciably lower ($\sim$10\%) and the central 
clusters is significantly higher at $\sim$26\%.   
The OB star candidates are distributed throughout the extended cluster region, c.f. Fig.~\ref{obvar}, 
with 20/29 not located in one of the subclusters.  Each of the four subclusters contains at least one 
massive star candidate (the central cluster contains at least 35: four new candidates in this study and 
31 from \citet{wol}); indicating that star formation may have been initiated independently in each of the 
subcluster regions, with a subsequent generation (possibly due to triggering) in the central region 
around IRS~2.  

The median $K$-band extinctions measured towards each subcluster and the distributed population 
are similar to within 1$\sigma$, consistent with all of the regions being at the same distance.  
Likewise, the median IRAC SED slope, $_{dered}\alpha_{IRAC}$, are similar. This suggests a similar 
age and evolutionary history across the whole region. This further supports the idea that star formation 
is not propagating through the region, but rather that the molecular cloud has undergone collapse and 
is forming stars in fractal subclusters. Further spectroscopic observations may yet find a trend towards 
older ages in the distributed population similar to that observed in Serpens and NGC~1333 \citep{win09}.   
The higher protostar fraction in the central cluster, rather than implying that the outer subclusters and 
distributed population are older per se, may be due to a second generation of protostars formed via 
triggering due to the IRS~2 O5.5 binary.  

The north west subcluster of YSOs corresponds to the 2MASS cluster Obj~36 \citep{dut}, which 
is associated with the reflection nebula vdBH-RN43 \citep{van} located at $08^h58^m04^s$, 
$-47^d22^m50^s$ in the region of RCW 38.  The association of this cluster with the main RCW 38 
cluster has not previously been established.  However, from the {\it Spitzer} observations, the subcluster 
does not appear to be physically isolated from the RCW~38 core; a band of YSOs stretches between the 
two.  Visually, the structure of the molecular cloud appears contiguous between the two regions, c.f. Fig.~\ref{field}.    
The $^{13}CO$ observations of \citet{yama} also show emission extending between the regions.      
The K-band extinctions, $A_K$, and magnitudes at J-band and 3.6${\mu}m$ were compared 
to those of the core and other subclusters; no difference was found in either the range or median of 
the extinction or magnitudes between the NW subcluster and the rest of the region.  This indicates 
that Obj~36 lies at a similar distance to the RCW~38 core and that it can be considered a subcluster 
in the extended RCW~38 star formation region.

\section{\bf Conclusion}

We have undertaken an extensive survey of the RCW~38 massive star forming region utilizing the 
capabilities of the {\it Spitzer} and {\it Chandra} space telescopes.  Combining mid-IR and X-ray 
observations of this young star forming region, with near-IR data from 2MASS, has enabled the 
identification of YSOs from the protostellar and disk-bearing phases to diskless young members.

\begin{itemize}

\item We identify 624 young stellar objects in RCW 38 in the overlap regions of our four epochs of IRAC 
data.  The initial identification process identified 712 objects, of which 32 were found to be contaminants. 
Of the remaining 680 YSOs, 56 lay outside the IRAC 4 band overlap region and were 
not included in the catalogue.   Of the 624 YSOs, 226 were detected in X-rays with {\it Chandra}.  The YSOs 
were of the following evolutionary classes:  23 class I, 90 flat spectrum, 437 class II, and 74  class III.

\item A further 177 candidate YSOs were identified using the [3.6] vs. [3.6-5.8] color-magnitude diagram.  
Sources with $8.0 \le m_{3.6} \le 14.0$ and a color more than $1~\sigma$ greater than $[3.6 - 5.8] > 0.4$ 
were selected.    As the contamination removal methods  could not be utilized  \citep{gut09}, we 
consider these to be tentative candidates.

\item Of the 433 YSOs located in the $Chandra$ field, 226 had confirmed X-ray detections. Nine of these 
were class 0/I protostars (9/22 or 41\% of the class 0/I in the {\it Chandra} fov), eighteen were flat spectrum 
(18/58 or 31\% of the flat spectrum sources), 125 were class II detections (125/279 or 45\%). 
The remaining 74 sources did not exhibit excess emission and were identified as class III members of the 
cluster;  74/624 or 12\% of the members were identified solely by {\it Chandra}.

\item The disk fraction, the ratio of the number of X-ray detected disk-bearing pre-main sequence stars to the 
total number of X-ray detected pre-main sequence stars,  of the YSOs with $mag_{3.6} > 12$ was found to be 
63$\pm$6\%, consistent with a cluster age of $\sim$1~Myrs.

\item We find an observed relationship of $N_H$$= 1.61\pm0.26\times10^{22}$$A_K$ from sixteen class II 
sources, and $N_H$$= 1.52\pm0.49\times10^{22}$$A_K$ from eight class III objects.
These values are consistent with the \citet{vuo} ratio of $1.6 \times 10^{22}$ for the local ISM and nearby 
molecular clouds,  and with the previous measurement of  $2.0 \times 10^{22}$ \citep{wol}.

\item By examining the photometric variance of the sources between epochs 72 variable candidates were 
identified.  Of these 48 had previously been identified as YSOs from their IR excess emission and X-ray 
luminosity.  The remaining 24 were found to consist of two class II sources and twenty-two candidate class IIIs. 
The percentage of identified variables by evolutionary class decreased through the class 0/I sources 
(30$\pm$11\%) to the class II (7$\pm$1\%).

\item An examination of the $K$ vs. $J - K$ color-magnitude diagram lead to the identification of 29 
candidate O-star members of the cluster, of which nine were also identified as variables.

\item Three young stars were identified by association with emission features at 24~$\mu$m; their 
IRAC colors and SEDs indicate that they are diskless class III members.   Another deeply embedded 
protostar was identified in the centre of RCW~38, slightly offset from the O5.5 binary.

\item The spatial distribution of the cluster was found to be fractally subclustered, with a $Q$-parameter of 0.41. 
The MST of the cluster had a characteristic branch length of 0.412~pc. An extended filament of protostars in the 
central subcluster appears to remain embedded in dense dust and gas, which may  trace the edge of density 
enhancement perhaps due to a shock front, and indicates ongoing and perhaps triggered star formation in the 
central core of RCW~38.

\item We identify four subclusters in RCW~38;  the central region of star formation in RCW~38 is surrounded 
by three subclusters and distributed young stars. Variable sources were identified in each of the four subclusters. 
The O star candidates are distributed throughout the cluster;  candidates are located in each of the four 
subclusters, while twenty candidates form part of the distributed population.   
The central cluster and distributed stars have higher protostellar fractions than the three subclusters.

\item The NW subcluster contains 62 members and is identified with a previously identified near-IR star cluster: 
Obj~36 \citep{dut}. Cloud morphology, continuous YSO distribution, and extinction measurements indicate that 
this region is not separate from, but is rather a subcluster of, RCW~38.

\end{itemize}

The authors would like to thank the referee for many helpful comments.   
This work is based on observations made with the {\it Spitzer} Space Telescope (PID 20127), 
which is operated by the Jet Propulsion Laboratory, California Institute of Technology under NASA 
contract 1407. Support for this work was provided by NASA through contract 1256790 issued by 
JPL/Caltech. Support for the IRAC instrument was provided by NASA through contract 960541 issued 
by JPL.
This publication makes use of data products from the Two Micron All Sky Survey, which is a 
joint project of the University of Massachusetts and the Infrared Processing and Analysis 
Center/California Institute of Technology, funded by the National Aeronautics and Space 
Administration and the National Science Foundation.
This research has made use of the NASA/IPAC Infrared Science Archive, which is operated by 
the Jet Propulsion Laboratory, California Institute of Technology, under contract with the 
National Aeronautics and Space Administration.
This research made use of Montage, funded by the National Aeronautics and Space Administration's 
Earth Science Technology Office, Computation Technologies Project, under Cooperative Agreement 
Number NCC5-626 between NASA and the California Institute of Technology. Montage is maintained 
by the NASA/IPAC Infrared Science Archive.
TLB acknowledges support from NASA through a grant for HST program 11123 from the Space Telescope 
Science Institute, which is operated by the Association of Universities for Research in Astronomy, 
Incorporated, under NASA contract NAS5-26555.



\clearpage

\center
\begin{deluxetable}{rrrrrr}
\tablecolumns{6}
\tabletypesize{\scriptsize}
\tablewidth{0pt}
\tablecaption{Summary of IRAC Epochs.  \protect\label{epochs}}
\tablehead{\colhead{Epoch ID}   & \colhead{Date}  & \colhead{Int. Time (s)}   & \colhead{RA (J2000)}   & \colhead{Dec (J2000)} &   \colhead{Coverage (arcmin)} }
\startdata

1  & 2006.06.05\/05:53:33  &  12  &  8:58:59.7  & -47:26:37.6  &  $\sim30\times30$ \\

2  & 2006.06.05\/06:58:53  &    2  &  8:59:25.9  &  -47:32:34.8  &  $\sim15\times15$  \\

3  & 2007.02.21\/19:42:03  &   12 &  8:59:05.7  & -47:30:43.6  & $\sim30\times30$  \\

4  & 2007.02.21\/20:45:35  &   2    &  8:59:05.9  &  -47:30:38.7 &  $\sim15\times15$ \\

\enddata

\end{deluxetable}
\endcenter

\clearpage

\center
\begin{deluxetable}{cccrrrrccr}
\tablecolumns{10}
\tabletypesize{\scriptsize}
\tablewidth{0pt}
\tablecaption{NGC 1333: {\it Chandra} X-ray Source Photometry  \protect\label{newxrays}}
\tablehead{\colhead{{\it Chandra}}  & \colhead{RA}   & \colhead{Dec} & \colhead{Raw}  &   \colhead{Net\,\tablenotemark{a}}  &   \colhead{$N_H$}   &   \colhead{kT}    &   \colhead{Abs. Flux}  &   \colhead{Tot. UnAbs. Flux}  &   \colhead{$\chi^{2}$}  \\
\colhead{ID}  & \colhead{(2000)}   & \colhead{(2000)}   & \colhead{CNT}  &   \colhead{CNT}  &   \colhead{$10^{22}cm^{-2}$}  &   \colhead{keV}  &   \multicolumn{2}{c}{$[erg cm^{-2} s^{-1}]$}   &   \colhead{} 
}

\startdata

1   &   8:59:55.58   &   -47:28:28.80   &   12.0   &   -0.3   &  \nodata    &   \nodata    &   \nodata   &  \nodata    &   \nodata    \\
2   &   8:59:53.60   &   -47:28:46.31   &   93.0   &   89.5   &   1.87$\pm$0.36   &   3.90$\pm$1.54   &   1.45E-14   &   3.28E-14   &   0.53    \\
3   &   8:59:52.33   &   -47:34:01.49   &   15.0   &   3.2   &   0.00$\pm$0.17   &   1.83$\pm$36.36   &   1.14E-15   &   1.14E-15   &   0.60    \\
4   &   8:59:50.36   &   -47:26:57.03   &   68.0   &   63.5   &   0.35$\pm$0.17   &   3.12$\pm$2.14   &   6.54E-15   &   1.02E-14   &   0.57    \\
5   &   8:59:47.99   &   -47:28:56.23   &   7.0   &   3.4   &   0.00$\pm$0.27   &   2.32$\pm$\nodata   &   4.18E-16   &   4.18E-16   &   0.37    \\
6   &   8:59:46.70   &   -47:31:53.62   &   71.0   &   67.0   &   0.75$\pm$0.25   &   4.63$\pm$5.88   &   8.52E-15   &   1.42E-14   &   0.57    \\
7   &   8:59:42.96   &   -47:31:58.63   &   38.0   &   34.1   &   1.90$\pm$0.81   &   7.54$\pm$\nodata   &   6.32E-15   &   1.18E-14   &   0.54    \\
8   &   8:59:41.58   &   -47:35:07.91   &   177.0   &   164.2   &   0.81$\pm$0.15   &   3.02$\pm$0.85   &   1.86E-14   &   3.63E-14   &   0.87    \\
\enddata

\tablenotetext{a}{Net Counts: background subtracted and aperture corrected from raw counts.  }
\tablenotetext{b}{ Abs. Flux: Absorbed flux. Tot. UnAbs. Flux: Total unabsorbed flux.  }

\end{deluxetable}
\endcenter

\clearpage

\center
\begin{deluxetable}{rrrrrr}
\tablecolumns{6}
\tabletypesize{\scriptsize}
\tablewidth{0pt}
\tablecaption{Astrometry of the Class 0/I YSOs in RCW 38.  \protect\label{tableastro1}}
\tablehead{\colhead{{\it Spitzer} ID}   & \colhead{Epoch ID}  & \colhead{RA (J2000)}   & \colhead{Dec (J2000)}  & \colhead{Class}  &   \colhead{X-ray, Variable\,\tablenotemark{a}} }
\startdata

72  &  3  &  8:59:05.86  &  -47:36:29.41  &  0/I  &  \nodata  \\

114  &  3  &  8:58:59.19  &  -47:32:32.86    &  0/I  &  x  \\

138  &  3  &  8:59:11.11  &  -47:31:23.85   &  0/I  &  x  \\

141  &  3  &  8:59:00.64  &  -47:27:24.65    &  0/I  &  x,v  \\

165  &  3  &  8:58:25.45  &  -47:31:34.96    &  0/I  &  v  \\

185  &  3  &  8:59:16.56  &  -47:30:25.30    &  0/I  &  \nodata  \\

239  &  3  &  8:59:27.93  &  -47:32:56.47    &  0/I  &  \nodata  \\

258  &  3  &  8:59:09.31  &  -47:29:18.92   &  0/I  &  x  \\

270  &  3  &  8:58:25.59  &  -47:30:51.43   &  0/I  &  v  \\

274  &  1  &  8:59:10.63  &  -47:22:39.05    &  0/I  &  \nodata  \\

291  &  3  &  8:59:15.30  &  -47:27:05.79   &  0/I  &  x,v  \\

292  &  3  &  8:58:39.39  &  -47:21:11.58    &  0/I  &  \nodata  \\

453  &  1  &  8:59:48.22  &  -47:17:57.92    &  0/I  &  \nodata  \\

472  &  4  &  8:59:01.28  &  -47:26:05.38    &  0/I  &  \nodata  \\

475  &  4  &  8:59:09.97  &  -47:26:09.38    &  0/I  &  \nodata  \\

478  &  3  &  8:59:13.78  &  -47:29:05.75    &  0/I  &  \nodata  \\

492  &  3  &  8:59:05.95  &  -47:31:51.48    &  0/I  &  x  \\

493  &  3  &  8:59:15.68  &  -47:30:9.78  &  0/I  &  x  \\

559  &  3  &  8:58:23.85  &  -47:32:40.80  &  0/I  &  v  \\

571  &  3  &  8:58:46.03  &  -47:34:47.68  &  0/I  &  v   \\

577  &  3  &  8:59:06.29  &  -47:31:07.65    &  0/I  &  x \\

581  &  3  &  8:59:02.92  &  -47:30:54.34    &  0/I  &  \nodata  \\

589  &  3  &  8:59:03.00  &  -47:30:44.50  &  0/I  &  x,v \\

\enddata

\tablenotetext{a}{v: indicates source is a candidate variable. }

\end{deluxetable}
\endcenter

\clearpage

\center
\begin{deluxetable}{rrrrrr}
\tablecolumns{6}
\tabletypesize{\scriptsize}
\tablewidth{0pt}
\tablecaption{Astrometry of the flat spectrum YSOs in RCW 38.  \protect\label{tableastrof}}
\tablehead{\colhead{{\it Spitzer} ID}   & \colhead{Epoch ID}  & \colhead{RA (J2000)}   & \colhead{Dec (J2000)}   & \colhead{Class}  &   \colhead{X-ray,Variable\,\tablenotemark{a}} }
\startdata

19  &  2  &  8:58:39.71  &  -47:36:58.28   &  FS  &  \nodata  \\

26  &  4  &  8:58:41.82  &  -47:38:22.66    &  FS  &  \nodata  \\

42  &  2  &  8:58:25.91  &  -47:36:34.24    &  FS  &  \nodata  \\

52  &  3  &  8:59:30.12  &  -47:38:39.42    &  FS  &  \nodata  \\

64  &  2  &  8:58:45.67  &  -47:38:24.67    &  FS  &  \nodata  \\

69  &  2  &  8:59:37.74  &  -47:38:43.84    &  FS  &  \nodata  \\

78  &  2  &  8:58:35.16  &  -47:40:09.08   &  FS  &  \nodata  \\

80  &  4  &  8:58:29.70  &  -47:38:18.43    &  FS  &  \nodata  \\
\enddata

\tablenotetext{a}{x: indicates Chandra X-ray detection.  v: indicates source is a candidate variable. }

\end{deluxetable}
\endcenter

\clearpage

\center
\begin{deluxetable}{rrrrrr}
\tablecolumns{6}
\tabletypesize{\scriptsize}
\tablewidth{0pt}
\tablecaption{Astrometry of the Class II YSOs in RCW 38.  \protect\label{tableastro2}}
\tablehead{\colhead{{\it Spitzer} ID}   & \colhead{Epoch ID}  & \colhead{RA (J2000)}   & \colhead{Dec (J2000)}   & \colhead{Class}  &   \colhead{X-ray, Var., O?\,\tablenotemark{a}} }
\startdata
1  &  4  &  8:58:40.06  &  -47:37:49.64  &    II  &  \nodata  \\

2  &  3  &  8:59:08.45  &  -47:37:19.47  &    II  & x  \\

3  &  3  &  8:59:41.57  &  -47:35:07.99  &    II  &  x  \\

4  &  4  &  8:57:59.98  &  -47:36:02.43  &    II  &  \nodata  \\

5  &  2  &  8:59:31.08  &  -47:41:36.04  &    II  &  \nodata  \\

6  &  4  &  8:58:00.31  &  -47:36:47.36  &    II  &  \nodata  \\

7  &  3  &  8:58:42.75  &  -47:35:13.33  &   II  &  \nodata  \\

8  &  2  &  8:59:11.80  &  -47:40:43.99  &   II  &  \nodata  \\
\enddata

\tablenotetext{a}{x: indicates Chandra X-ray detection.  v: indicates source is a candidate variable. o: indicates source is an O-star candidate. }

\end{deluxetable}
\endcenter

\clearpage

\center
\begin{deluxetable}{rrrrrr}
\tablecolumns{6}
\tabletypesize{\scriptsize}
\tablewidth{0pt}
\tablecaption{ Astrometry of the Class III YSOs in RCW 38. \protect\label{tableastro3}}
\tablehead{\colhead{{\it Spitzer} ID}   & \colhead{Epoch ID}  & \colhead{RA (J2000)}   & \colhead{Dec (J2000)}   & \colhead{Class}  &   \colhead{X-ray/Var\,\tablenotemark{a}} }
\startdata
17  &  3  &  8:59:37.41  &  -47:34:35.17  &    III  &  x  \\

36  &  3  &  8:59:09.13  &  -47:34:05.24  &    III  &  x  \\

65  &  3  &  8:59:25.72  &  -47:34:36.20  &    III  &  x  \\

77  &  3  &  8:59:27.72  &  -47:35:50.73  &    III  &  x  \\

82  &  3  &  8:59:08.84  &  -47:36:37.77  &    III  &  x  \\

93  &  1  &  8:58:48.81  &  -47:32:15.96  &    III  &  x  \\

99  &  3  &  8:58:43.51  &  -47:29:34.32  &    III  &  x  \\

100  &  1  &  8:59:46.88  &  -47:38:04.42  &    III  &  x  \\
\enddata

\tablenotetext{a}{x: indicates source is a Chandra X-ray detection. }

\end{deluxetable}
\endcenter

\clearpage

\center
\begin{deluxetable}{rrrrrrrrrrr}
\tablecolumns{11}
\tabletypesize{\scriptsize}
\setlength{\tabcolsep}{0.03in}
\tablewidth{0pt}
\tablecaption{Photometry of the Class 0/I YSOs in RCW 38. \protect\label{tablephot1}}
\tablehead{  \colhead{{\it Sp.}} & \colhead{Ep.} & \colhead{J} & \colhead{H}  & \colhead{K}   & \colhead{3.6${\mu}m$}  & \colhead{4.5${\mu}m$}  &   \colhead{5.8${\mu}m$}  &  \colhead{8.0${\mu}m$}   &  \colhead{$\alpha_{IRAC}$}   &  \colhead{$A_{K}$}  \\
\colhead{ID}  &  \colhead{ID}  &  \colhead{[mag.]}  &  \colhead{[mag.]}  &  \colhead{[mag.]}  &  \colhead{[mag.]}   &  \colhead{[mag.]}  &  \colhead{[mag.]}  &  \colhead{[mag.]}  &  \colhead{}  &  \colhead{[mag.]}   }
\startdata
72  &  3  &  16.17$\pm$0.10  &  15.03$\pm$0.07  &  14.40$\pm$0.09  &  13.08$\pm$0.14  &  12.88$\pm$0.14  &  10.85$\pm$0.23  &  \nodata  &  1.34  &  0.41  \\

114  &  3  &  \nodata  &  \nodata  &  13.95$\pm$0.07  &  11.35$\pm$0.10  &  10.65$\pm$0.06  &  9.28$\pm$0.18  &  \nodata  &  1.13  &  \nodata  \\

138  &  3  &  \nodata  &  \nodata  &  15.00$\pm$0.16  &  11.59$\pm$0.14  &  10.68$\pm$0.13  &  \nodata  &  \nodata  &  0.71  &  \nodata  \\

141  &  3  &  14.66$\pm$0.05  &  13.10$\pm$0.03  &  \nodata  &  9.88$\pm$0.02  &  9.07$\pm$0.02  &  8.16$\pm$0.02  &  6.98$\pm$0.05  &  0.49  &  \nodata  \\

165  &  3  &  \nodata  &  \nodata  &  14.93$\pm$0.12  &  12.33$\pm$0.07  &  10.94$\pm$0.04  &  9.96$\pm$0.05  &  8.72$\pm$0.08  &  1.23  &  \nodata  \\

185  &  3  &  \nodata  &  \nodata  &  14.62$\pm$0.13  &  12.58$\pm$0.10  &  11.41$\pm$0.06  &  9.88$\pm$0.13  &  8.48$\pm$0.18  &  1.95  &  \nodata  \\

239  &  3  &  16.18$\pm$0.11  &  14.83$\pm$0.08  &  13.84$\pm$0.09  &  13.18$\pm$0.21  &  12.03$\pm$0.21  &  \nodata  &  \nodata  &  1.65  &  \nodata  \\

258  &  3  &  \nodata  &  \nodata  &  12.61$\pm$0.07  &  10.29$\pm$0.09  &  9.65$\pm$0.07  &  \nodata  &  \nodata  &  -0.35  &  \nodata  \\
\enddata

\end{deluxetable}
\endcenter

\center
\begin{deluxetable}{rrrrrrrrrrr}
\tablecolumns{11}
\tabletypesize{\scriptsize}
\setlength{\tabcolsep}{0.03in}
\tablewidth{0pt}
\tablecaption{Photometry of the flat spectrum YSOs in RCW 38. \protect\label{tablephotf}}
\tablehead{  \colhead{{\it Sp.}} & \colhead{Ep.} & \colhead{J} & \colhead{H}  & \colhead{K}   & \colhead{3.6${\mu}m$}  & \colhead{4.5${\mu}m$}  &   \colhead{5.8${\mu}m$}  &  \colhead{8.0${\mu}m$}   &  \colhead{$\alpha_{IRAC}$}   &  \colhead{$A_{K}$}  \\
\colhead{ID}  &  \colhead{ID}  &  \colhead{[mag.]}  &  \colhead{[mag.]}  &  \colhead{[mag.]}  &  \colhead{[mag.]}   &  \colhead{[mag.]}  &  \colhead{[mag.]}  &  \colhead{[mag.]}  &  \colhead{}  &  \colhead{[mag.]}   }
\startdata
19  &  II  &  \nodata  &  15.86$\pm$0.13  &  14.83$\pm$0.12  &  13.60$\pm$0.12  &  13.22$\pm$0.12  &  12.48$\pm$0.13  &  11.45$\pm$0.16  &  -0.35  &  \nodata  \\

26  &  4  &  \nodata  &  15.66$\pm$0.13  &  14.14$\pm$0.07  &  \nodata  &  11.97$\pm$0.07  &  11.31$\pm$0.08  &  10.47$\pm$0.13  &  -0.43  &  \nodata  \\

42  &  II  &  \nodata  &  16.32$\pm$0.18  &  15.12$\pm$0.14  &  13.93$\pm$0.14  &  13.35$\pm$0.13  &  12.88$\pm$0.16  &  11.71$\pm$0.19  &  -0.36  &  \nodata  \\

52  &  3  &  \nodata  &  15.41$\pm$0.12  &  14.52$\pm$0.12  &  13.18$\pm$0.10  &  12.51$\pm$0.09  &  11.79$\pm$0.08  &  10.91$\pm$0.09  &  -0.24  &  \nodata  \\

64  &  II  &  \nodata  &  15.64$\pm$0.12  &  14.44$\pm$0.09  &  13.58$\pm$0.12  &  13.03$\pm$0.11  &  12.19$\pm$0.11  &  10.80$\pm$0.10  &  0.37  &  \nodata  \\

69  &  2  &  16.72$\pm$0.18  &  15.13$\pm$0.08  &  14.16$\pm$0.07  &  12.80$\pm$0.08  &  12.18$\pm$0.07  &  11.52$\pm$0.10  &  10.63$\pm$0.19  &  -0.38  &  0.08  \\

78  &  2  &  16.37$\pm$0.11  &  14.86$\pm$0.05  &  14.17$\pm$0.06  &  13.73$\pm$0.13  &  13.16$\pm$0.12  &  12.68$\pm$0.15  &  11.36$\pm$0.17  &  -0.19  &  \nodata  \\

80  &  4  &  \nodata  &  15.73$\pm$0.14  &  14.82$\pm$0.13  &  12.82$\pm$0.08  &  12.03$\pm$0.07  &  11.38$\pm$0.07  &  10.25$\pm$0.07  &  0.05  &  \nodata  \\
\enddata

\end{deluxetable}
\endcenter

\center
\begin{deluxetable}{rrrrrrrrrrr}
\tablecolumns{11}
\tabletypesize{\scriptsize}
\setlength{\tabcolsep}{0.03in}
\tablewidth{0pt}
\tablecaption{Photometry of the Class II YSOs in RCW 38. \protect\label{tablephot2}}
\tablehead{  \colhead{{\it Sp.}} & \colhead{Ep.} & \colhead{J} & \colhead{H}  & \colhead{K}   & \colhead{3.6${\mu}m$}  & \colhead{4.5${\mu}m$}  &   \colhead{5.8${\mu}m$}  &  \colhead{8.0${\mu}m$}   &  \colhead{$\alpha_{IRAC}$}   &  \colhead{$A_{K}$}  \\
\colhead{ID}  &  \colhead{ID}  &  \colhead{[mag.]}  &  \colhead{[mag.]}  &  \colhead{[mag.]}  &  \colhead{[mag.]}   &  \colhead{[mag.]}  &  \colhead{[mag.]}  &  \colhead{[mag.]}  &  \colhead{}  &  \colhead{[mag.]}   }
\startdata
1  &  4  &  \nodata  &  15.49$\pm$0.11  &  14.63$\pm$0.09  &  13.46$\pm$0.11  &  13.02$\pm$0.11  &  12.54$\pm$0.13  &  11.68$\pm$0.16  &  -0.84  &  \nodata  \\

2  &  3  &  15.59$\pm$0.07  &  14.10$\pm$0.04  &  13.49$\pm$0.05  &  12.99$\pm$0.09  &  12.95$\pm$0.11  &  \nodata  &  \nodata  &  -3.08  &  0.94  \\

3  &  3  &  15.55$\pm$0.07  &  13.88$\pm$0.05  &  13.01$\pm$0.04  &  11.89$\pm$0.06  &  11.49$\pm$0.05  &  11.28$\pm$0.07  &  10.74$\pm$0.12  &  -1.75  &  1.01  \\

4  &  4  &  \nodata  &  15.46$\pm$0.14  &  14.39$\pm$0.08  &  \nodata  &  12.31$\pm$0.08  &  \nodata  &  11.35$\pm$0.24  &  -1.32  &  \nodata  \\

5  &  2  &  16.93$\pm$0.22  &  15.73$\pm$0.14  &  14.97$\pm$0.14  &  13.51$\pm$0.12  &  13.00$\pm$0.11  &  12.72$\pm$0.14  &  12.45$\pm$0.36  &  -1.69  &  \nodata  \\

6  &  4  &  16.98$\pm$0.20  &  15.13$\pm$0.08  &  14.28$\pm$0.08  &  \nodata  &  12.28$\pm$0.08  &  \nodata  &  10.98$\pm$0.13  &  -0.75  &  \nodata  \\

7  &  3  &  16.78$\pm$0.15  &  14.26$\pm$0.03  &  12.71$\pm$0.04  &  10.93$\pm$0.04  &  10.25$\pm$0.03  &  9.71$\pm$0.03  &  8.88$\pm$0.05  &  -0.74  &  1.31  \\

8  &  2  &  15.86$\pm$0.09  &  14.51$\pm$0.05  &  13.82$\pm$0.07  &  12.89$\pm$0.09  &  12.49$\pm$0.09  &  12.14$\pm$0.13  &  11.06$\pm$0.19  &  -0.91  &  0.67  \\
\enddata

\end{deluxetable}
\endcenter

\center
\begin{deluxetable}{rrrrrrrrrrr}
\tablecolumns{11}
\tabletypesize{\scriptsize}
\setlength{\tabcolsep}{0.03in}
\tablewidth{0pt}
\tablecaption{Photometry of the Class III YSOs in RCW 38. \protect\label{tablephot3}}
\tablehead{  \colhead{{\it Sp.}} & \colhead{Ep.} & \colhead{J} & \colhead{H}  & \colhead{K}   & \colhead{3.6${\mu}m$}  & \colhead{4.5${\mu}m$}  &   \colhead{5.8${\mu}m$}  &  \colhead{8.0${\mu}m$}   &  \colhead{$\alpha_{IRAC}$}   &  \colhead{$A_{K}$}  \\
\colhead{ID}  &  \colhead{ID}  &  \colhead{[mag.]}  &  \colhead{[mag.]}  &  \colhead{[mag.]}  &  \colhead{[mag.]}   &  \colhead{[mag.]}  &  \colhead{[mag.]}  &  \colhead{[mag.]}  &  \colhead{}  &  \colhead{[mag.]}   }

\startdata

17  &  3  &  15.85$\pm$0.10  &  14.71$\pm$0.07  &  14.20$\pm$0.08  &  13.83$\pm$0.14  &  13.72$\pm$0.15  &  \nodata  &  \nodata  &  -2.67  &  0.52  \\

36  &  3  &  14.94$\pm$0.04  &  13.67$\pm$0.03  &  13.38$\pm$0.03  &  12.83$\pm$0.10  &  13.17$\pm$0.24  &  \nodata  &  \nodata  &  -4.48  &  0.71  \\

65  &  3  &  12.22$\pm$0.02  &  11.97$\pm$0.03  &  11.76$\pm$0.02  &  11.75$\pm$0.06  &  11.70$\pm$0.07  &  \nodata  &  \nodata  &  -2.66  &  \nodata  \\

77  &  3  &  14.19$\pm$0.03  &  13.42$\pm$0.03  &  13.04$\pm$0.03  &  12.74$\pm$0.08  &  12.69$\pm$0.09  &  12.68$\pm$0.15  &  \nodata  &  -2.78  &  0.05  \\

82  &  3  &  11.64$\pm$0.02  &  11.39$\pm$0.03  &  11.34$\pm$0.03  &  11.37$\pm$0.04  &  11.37$\pm$0.05  &  11.47$\pm$0.08  &  11.98$\pm$0.55  &  -3.61  &  \nodata  \\

93  &  1  &  15.31$\pm$0.07  &  14.33$\pm$0.09  &  13.87$\pm$0.07  &  \nodata  &  13.65$\pm$0.25  &  \nodata  &  \nodata  &  \nodata &  0.32  \\

99  &  3  &  16.56$\pm$0.16  &  14.88$\pm$0.06  &  14.05$\pm$0.07  &  13.19$\pm$0.12  &  13.27$\pm$0.14  &  \nodata  &  \nodata  &  -3.19  &  \nodata  \\

100  &  1  &  15.21$\pm$0.05  &  14.18$\pm$0.04  &  13.81$\pm$0.06  &  13.57$\pm$0.12  &  \nodata  &  13.53$\pm$0.27  &  \nodata  &  -2.94  &  0.45  \\
\enddata

\end{deluxetable}
\endcenter

\center
\begin{deluxetable}{rrrrrrrrrrr}
\tablecolumns{11}
\tabletypesize{\scriptsize}
\setlength{\tabcolsep}{0.03in}
\tablewidth{0pt}
\tablecaption{Photometry of the Candidate [3.6] v. [3.6-5.8] YSOs in RCW 38. \protect\label{tablephotcandy}}
\tablehead{  \colhead{{\it Sp.}} & \colhead{Ep.} & \colhead{RA}  & \colhead{Dec} & \colhead{J} & \colhead{H}  & \colhead{K}   & \colhead{3.6${\mu}m$}  & \colhead{4.5${\mu}m$}  &   \colhead{5.8${\mu}m$}  &  \colhead{8.0${\mu}m$}   \\
\colhead{ID}  &  \colhead{ID}   &  \colhead{J2000}   &  \colhead{J2000}  &  \colhead{[mag.]}  &  \colhead{[mag.]}  &  \colhead{[mag.]}  &  \colhead{[mag.]}   &  \colhead{[mag.]}  &  \colhead{[mag.]}  &  \colhead{[mag.]}   }

\startdata

605  &  4  &  8:59:31.23  &  -47:35:21.43  &  16.50$\pm$0.15  &  15.53$\pm$0.12  &  14.77$\pm$0.12  &  14.31$\pm$0.17  &  14.17$\pm$0.19  &  12.87$\pm$0.16  &  \nodata  \\

606  &  4  &  8:59:34.90  &  -47:34:27.35  &  15.99$\pm$0.09  &  14.39$\pm$0.05  &  13.70$\pm$0.05  &  13.21$\pm$0.10  &  13.13$\pm$0.12  &  12.53$\pm$0.14  &  \nodata  \\

607  &  4  &  8:58:42.74  &  -47:36:43.69  &  16.72$\pm$0.15  &  14.73$\pm$0.06  &  14.18$\pm$0.07  &  13.52$\pm$0.12  &  13.38$\pm$0.13  &  13.03$\pm$0.19  &  \nodata  \\

609  &  4  &  8:59:33.28  &  -47:37:8.86  &  16.65$\pm$0.17  &  15.34$\pm$0.11  &  14.76$\pm$0.11  &  13.82$\pm$0.14  &  13.53$\pm$0.14  &  13.01$\pm$0.22  &  12.44$\pm$0.35  \\

610  &  4  &  8:58:27.16  &  -47:34:48.51  &  16.40$\pm$0.11  &  14.83$\pm$0.06  &  14.41$\pm$0.07  &  13.75$\pm$0.13  &  13.62$\pm$0.15  &  12.63$\pm$0.16  &  \nodata  \\

611  &  2  &  8:59:13.96  &  -47:41:39.99  &  15.59$\pm$0.06  &  14.14$\pm$0.04  &  13.57$\pm$0.05  &  13.16$\pm$0.10  &  13.17$\pm$0.12  &  12.53$\pm$0.16  &  \nodata  \\

612  &  4  &  8:58:40.20  &  -47:36:13.47  &  16.78$\pm$0.17  &  14.90$\pm$0.07  &  14.16$\pm$0.06  &  13.50$\pm$0.12  &  13.33$\pm$0.13  &  12.22$\pm$0.12  &  \nodata  \\

613  &  4  &  8:58:22.97  &  -47:34:43.75  &  \nodata  &  13.53$\pm$0.05  &  12.92$\pm$0.04  &  12.54$\pm$0.08  &  12.47$\pm$0.09  &  11.91$\pm$0.14  &  \nodata  \\
\enddata

\end{deluxetable}
\endcenter

\center
\begin{deluxetable}{rrrrrrrrrrrrr}
\tablecolumns{13}
\rotate
\tabletypesize{\scriptsize}
\setlength{\tabcolsep}{0.03in}
\tablewidth{0pt}
\tablecaption{Astrometry \& Photometry of Candidate Variable YSOs in RCW 38. \protect\label{tablevariables}}
\tablehead{  \colhead{{\it Sp.}} & \colhead{Ep.} & \colhead{RA}  & \colhead{Dec}  & \colhead{J} & \colhead{H}  & \colhead{K}   & \colhead{3.6${\mu}m$}  & \colhead{4.5${\mu}m$}  &  \colhead{5.8${\mu}m$} &  \colhead{8.0${\mu}m$}   &  \colhead{$A_{K}$} & \colhead{Class} \\
\colhead{ID}  &  \colhead{ID}  &  \colhead{J2000}   &  \colhead{J2000}  &  \colhead{[mag.]}  &  \colhead{[mag.]}  &  \colhead{[mag.]}  &  \colhead{[mag.]}   &  \colhead{[mag.]}  &  \colhead{[mag.]}  &  \colhead{[mag.]}    &  \colhead{[mag.]}  & \colhead{} }

\startdata 

    oc1-v1 		& 2 	&        8:58:28.56 &      -47:39:00.9 &       9.991$\pm$0.026  &       8.205$\pm$0.044 &       7.413$\pm$0.026 &      \nodata   &      \nodata    &       6.855$\pm$0.008 &      6.813$\pm$0.012   &  1.25  &  III    \\      
    oc2-v2 		&  3 	&        8:58:59.40 &      -47:34:16.6 &       8.654$\pm$0.021  &       7.784$\pm$0.039 &       7.489$\pm$0.029 &      7.385$\pm$0.008 &       7.471$\pm$0.009 &      7.365$\pm$0.063  &       7.234$\pm$0.257   &    0.28  &  III       \\
    v3 			& 2 	&        8:58:53.32 &      -47:41:02.9 &       11.533$\pm$0.023  &       10.446$\pm$0.027 &       9.989$\pm$0.022 &       9.784$\pm$0.021 &      9.793$\pm$0.025 &       9.747$\pm$0.0319 &      9.660$\pm$0.045   &  0.48  &  III         \\
    v4 			& 2 	&        8:59:44.28 &      -47:42:44.9 &       11.723$\pm$0.021  &       10.379$\pm$0.025 &       9.903$\pm$0.024 &       9.603$\pm$0.019 &      9.664$\pm$0.023 &       9.533$\pm$0.029 &     9.553$\pm$0.048   &   0.78   &  III       \\
    v5 			& 1 	&        8:59:55.54 &      -47:40:48.4 &       11.615$\pm$0.021 &       10.267$\pm$0.024 &       9.763$\pm$0.024 &       9.516$\pm$0.018 &      9.326$\pm$0.020 &       9.215$\pm$0.024 &      9.192$\pm$0.038   &   0.79   &  III       \\       
    oc5-v6 		& 1 	&        8:58:20.07 &      -47:23:48.6 &       9.094$\pm$0.030 &       8.372$\pm$0.049 &       8.012$\pm$0.038 &      \nodata &       8.110$\pm$0.012   &       7.451$\pm$0.011 &       7.249$\pm$0.018   &    \nodata     &  III     \\    
    oc6-v7 		& 1 	&        8:57:46.59 &      -47:32:32.4 &       9.793$\pm$0.021 &       8.425$\pm$0.039 &       7.710$\pm$0.036 &      \nodata &      \nodata &       6.802$\pm$0.008 &       6.723$\pm$0.012   &     0.68   &  III     \\
\enddata

\end{deluxetable}
\endcenter

\center
\begin{deluxetable}{rrrrrrrrrrrr}
\tablecolumns{12}
\tabletypesize{\scriptsize}
\setlength{\tabcolsep}{0.03in}
\tablewidth{0pt}
\tablecaption{Astrometry Photometry of candidate O-type stars in RCW 38. \protect\label{tableOstars}}
\tablehead{  \colhead{{\it Sp.}} & \colhead{Ep.} & \colhead{RA}  & \colhead{Dec}  & \colhead{J} & \colhead{H}  & \colhead{K}   & \colhead{3.6${\mu}m$}  & \colhead{4.5${\mu}m$}  &   \colhead{5.8${\mu}m$}  &  \colhead{8.0${\mu}m$}    &  \colhead{$A_{K}$}  \\
\colhead{ID}  &  \colhead{ID}  &  \colhead{J2000}   &  \colhead{J2000}  &  \colhead{[mag.]}  &  \colhead{[mag.]}  &  \colhead{[mag.]}  &  \colhead{[mag.]}   &  \colhead{[mag.]}  &  \colhead{[mag.]}  &  \colhead{[mag.]}    &  \colhead{[mag.]}   }

\startdata

oc1-v1		& 2  &      8:58:28.56   &   -47:39:00.9   &    9.99$\pm$0.026   &    8.204$\pm$0.044   &    7.413$\pm$0.026   &   \nodata   &   \nodata   &    6.855$\pm$0.008   &   6.813$\pm$0.012   &  1.25  \\
oc2-v2		& 3  &      8:58:59.40   &   -47:34:16.6   &    8.65$\pm$0.021   &    7.783$\pm$0.040    &    7.488$\pm$0.029   &    7.386$\pm$0.008    &   7.472$\pm$0.010     &   7.365$\pm$0.064    &   7.234$\pm$0.257   &  0.28  \\
oc3			& 2  &      8:59:34.59   &   -47:39:51.8   &    12.31$\pm$0.021   &    10.334$\pm$0.025  &    9.460$\pm$0.021   &    9.084$\pm$0.015    &   9.004$\pm$0.017     &    8.784$\pm$0.020   &   8.786$\pm$0.042   &  1.45  \\
604			& 2  &      8:58:42.91   &   -47:39:37.6   &    10.40$\pm$0.027   &    8.994$\pm$0.022   &    8.498$\pm$0.021   &    8.689$\pm$0.013    &   8.388$\pm$0.013     &    8.111$\pm$0.015   &   8.086$\pm$0.023   &  0.85  \\
oc5-v6		& 1  &      8:58:20.07   &   -47:23:48.6   &    9.09$\pm$0.030   &    8.372$\pm$0.049   &    8.012$\pm$0.038   &   \nodata   &   8.110$\pm$0.012     &  7.451$\pm$0.011     &   7.249$\pm$0.018   &  \nodata \\
oc6-v7		& 1  &      8:57:46.59   &   -47:32:32.4   &    9.79$\pm$0.021   &    8.425$\pm$0.040   &    7.710$\pm$0.036   &   \nodata   &   \nodata   &    6.802$\pm$ 0.008   &   6.723$\pm$0.012   &  0.69  \\
oc7-v10		& 2  &      8:58:15.87   &   -47:30:18.8   &    11.92$\pm$0.024   &    9.854$\pm$0.022   &    8.975$\pm$0.021   &    8.859$\pm$0.014   &    8.537$\pm$0.014     &  8.314$\pm$0.017     &   8.265$\pm$0.025   &  1.54  \\
oc8			& 1  &      8:58:9.835   &   -47:13:32.4   &    11.02$\pm$0.026   &    9.321$\pm$0.022   &    8.585$\pm$0.021   &   \nodata   &   8.429$\pm$0.013     &  8.095$\pm$0.015     &   8.101$\pm$0.026   &  1.16 \\
\enddata

\end{deluxetable}
\endcenter

\center
\begin{deluxetable}{rrrrrrrrrrrr}
\tablecolumns{12}
\tabletypesize{\scriptsize}
\setlength{\tabcolsep}{0.03in}
\tablewidth{0pt}
\tablecaption{Astrometry \& Photometry of  stars associated with 24~${\mu}m$ emission in RCW 38. \protect\label{table24}}
\tablehead{  \colhead{{\it Sp.}} & \colhead{Ep.} & \colhead{RA}  & \colhead{Dec}  & \colhead{J} & \colhead{H}  & \colhead{K}   & \colhead{3.6${\mu}m$}  & \colhead{4.5${\mu}m$}  &   \colhead{5.8${\mu}m$}  &  \colhead{8.0${\mu}m$}    &  \colhead{$A_{K}$}  \\
\colhead{ID}  &  \colhead{ID}  &  \colhead{J2000}   &  \colhead{J2000}  &  \colhead{[mag.]}  &  \colhead{[mag.]}  &  \colhead{[mag.]}  &  \colhead{[mag.]}   &  \colhead{[mag.]}  &  \colhead{[mag.]}  &  \colhead{[mag.]}    &  \colhead{[mag.]}   }

\startdata

e1 &  3  &    -8:59:06.27   &   -47:24:31.0  &    12.030$\pm$0.022   &    10.938$\pm$0.026   &    10.345$\pm$0.023   &    10.016$\pm$0.025   &    9.848$\pm$0.030   &    9.641$\pm$0.108   &   8.561$\pm$0.108   &   0.36    	 	    \\
e2 &  3  &    -8:58:55.36   &   -47:23:24.8  &    10.398$\pm$0.024   &    9.8610$\pm$0.026   &    9.5430$\pm$0.023   &    9.3751$\pm$0.022   &    9.271$\pm$0.032   &    8.635$\pm$0.074   &   \nodata			&   \nodata	    \\
e3 &  3  &    -8:59:50.53   &   -47:24:31.3  &    9.6250$\pm$0.024   &    9.3570$\pm$0.033   &    9.0869$\pm$0.027   &    9.1790$\pm$0.015   &    9.081$\pm$0.018   &    9.046$\pm$0.028   &    8.875$\pm$0.07	&   \nodata	    \\
         
\enddata

\end{deluxetable}
\endcenter

\center
\begin{deluxetable}{rrrrrr}
\tablecolumns{6}
\tabletypesize{\scriptsize}
\setlength{\tabcolsep}{0.03in}
\tablewidth{0pt}
\tablecaption{K-S test probabilities of YSOs in RCW 38 by evolutionary class. \protect\label{tableprobs}}
\tablehead{ \colhead{} &  \colhead{class 0/I} & \colhead{flat spectrum} & \colhead{class II}  & \colhead{class III}  & \colhead{random}  }

\startdata 

{\b Class 0/I}    &     \nodata    &   0.31   &   $1.2\times10^{-5}$    &     0.016     &     $6.9\times10^{-3}$      \\

{\b Flat Spectrum}    &     0.31    &   \nodata   &   $1.3\times10^{-9}$    &     0.02     &     $5.5\times10^{-6}$      \\

{\b Class II}    &     $1.2\times10^{-5}$     &   $1.3\times10^{-9}$    &   \nodata    &     $2.6\times10^{-3}$      &     $2\times10^{-7}$      \\

{\b Class III}    &     0.016    &   0.02   &   $2.6\times10^{-3}$    &     \nodata     &     0.26     \\

\enddata

\end{deluxetable}
\endcenter

\center
\begin{deluxetable}{rrrrrrrrrrrr}
\tablecolumns{12}
\tabletypesize{\scriptsize}
\setlength{\tabcolsep}{0.03in}
\tablewidth{0pt}
\tablecaption{Properties of RCW~38 Subclusters. \protect\label{tablesubcs}}
\tablehead{ \colhead{cluster}  &  \colhead{total}   &  \colhead{C0/I\,\tablenotemark{a}} & \colhead{FS} & \colhead{CII}  & \colhead{CIII}  & \colhead{Var}  &  \colhead{OB}  &  \colhead{A$_K$}  &  \colhead{$^{dered}\alpha_{irac}$}  &  \colhead{Protostars (\%)}  &  \colhead{Variables (\%)} }

\startdata 

NE   &   11   &   1   &   1   &   9   &   \nodata   &   1   &   1   &   0.684$\pm$0.382   &   -0.996$\pm$0.802   &   18.2$\pm$12.8\%    &    9.1$\pm$9.1\% \\  

NW   &   62   &   \nodata   &   9   &   53   &   \nodata   &   8   &   1   &   0.608$\pm$0.225   &   -1.078$\pm$0.671    &   14.5$\pm$4.8\%    &    12.9$\pm$4.5\%   \\  

SW   &   41   &   \nodata   &   4   &   37   &   \nodata   &   3   &   1   &   0.530$\pm$0.244   &   -1.146$\pm$0.541    &   9.7$\pm$4.9\%    &    7.3$\pm$4.2\%   \\  

CN   &   255   &  15   &   39   &   154   &   47   &   24   &   6   &   0.836$\pm$0.747   &   -0.957$\pm$1.295    &   25.9$\pm$3.5\%    &    9.4$\pm$1.9\%   \\  

Dist.   &   222   &   7   &   31   &   157   &   27   &   36   &   20   &   0.701$\pm$0.448   &   -1.056$\pm$0.808    &   19.5$\pm$3.2\%    &    16.2$\pm$2.7\%   \\

\enddata

\tablenotetext{a}{C0/I: class 0/I,  FS: flat spectrum, CII: class II, CIII: class III, Var: variables, OB: OB star candidates, A$_K$: extinction at K-band, $^{dered}\alpha_{irac}$: slope of IRAC SED.   }

\end{deluxetable}
\endcenter


\clearpage

\clearpage

\clearpage

\clearpage

\clearpage

\clearpage

\clearpage

\clearpage

\clearpage

\clearpage

\clearpage

\clearpage

\clearpage

\clearpage

\clearpage

\clearpage


\clearpage


\begin{thebibliography}{}

\bibitem[Allen et al.(2004)]{all} Allen, L., Calvet, N., D'Alessio, P., et al, 2004, ApJS, 154, 363-366.

\bibitem[Andre \& Montmerle(1994)]{and2} Andre, P., Montmerle, T., 1994, ApJ, 420, 837.

\bibitem[Baraffe(1998)]{bar} Baraffe, I.,  1998, A\&A, 337, 403.

\bibitem[Blum et al.(2006)]{blum} Blum, R.D., Mould, J.R., Olsen, K.A., et al., 2006, AJ, 132, 2034-2045.   

\bibitem[Calvet et al.(1994)]{cal2} Calvet, N., Hartmann, L., Kenyon, S.J., Whitney, B.A., 1994, ApJ, 434, 330.

\bibitem[Cardelli et al.(1989)]{card} Cardelli, J., Clayton, G., Mathis, j., 1989, ApJ, 345, 245.

\bibitem[Carpenter et al.(2001)]{carp} Carpenter, J.M., Hillenbrand, L.A., Skrutskie, M.F., 2001, AJ, 121, 3160.  

\bibitem[Cartwright \& Whitworth(2004)]{car} Cartwright, A., Whitworth, A., 2004, MNRAS, 348, 589.

\bibitem[Casanova et al.(1995)]{cas} Casanova, S., Montmerle, T., Feigelson, E.D., Andre, P., 1995, ApJ, 439, 752.

\bibitem[Casertano \& Hut(1985)]{case} Casertano, S., Hut, P., 1985, ApJ, 298, 80. 

\bibitem[Comer\'{o}n et al.(2005)]{com} Comer\'{o}n, F., Schneider, N., Russeil, D., 2005, A\&A, 433, 955-977.  

\bibitem[Damiani et al.(1996)]{dam} Damiani, F., Maggio, A., Micela, G., \& Sciortino, S.\ 1996, Astronomical Data Analysis Software and Systems V, 101, 143 

\bibitem[D'Antona \& Mazzitelli(1997)]{dan} D'Antona, F., Mazzitelli, I., 1997, Mem. S.A.It., 68, 807.  

\bibitem[DeRose et al.(2009)]{der} DeRose, K. L., Bourke, T. L., Gutermuth, R. A., et al., 2009, AJ, 138, 33.

\bibitem[Doppman et al.(2005)]{dop} Doppman, G.W., Greene T.P., Covey K.R., Lada C.J., 2005, ApJ, 130, 1145-1170.

\bibitem[Dutra \& Bica(2001)]{dut} Dutra, C.M., Bica, E., 2001, A\&A, 376, 434-440.   

\bibitem[Eisenhardt et al.(2004)]{eis} Eisenhardt, P.R.,Stern, D., Brodwin, M., et al., 2004, ApJS, 154, 48. 

\bibitem[Evans et al.(2010)]{eva} Evans, I.~N., Primini, F.A., Glotfelty, K.J., Anderson, C.S., et al., 2010, \apjs, 189, 37 

\bibitem[Favata et al.(2003)]{fav1} Favata, Giardino, G., Micela, G., et al., 2003, A\&A, 403, 187.

\bibitem[Favata et al.(2005)]{fav2} Favata, F., Flaccomio, E., Reale, F., et al. , 2005, ApJS, 160, 469.

\bibitem[Fazio et al.(2004)]{faz} Fazio, G.G., Hora, J.L, Allen L.E., et al., 2004, ApJS, 154, 10.

\bibitem[Feigelson \& Townsley(2008)]{fei2008}   Feigelson, E., Townsley, L., 2008, ApJ, 673, 354.     

\bibitem[Feigelson et al.(2005)]{fei} Feigelson, E., Getman, R., Townsley, L., et al., 2005, ApJS, 160, 379.

\bibitem[Feigelson \& Getman(2005)]{feiget} Feigelson, E. \& Getman, R., 2005, astro-ph/0501207.

\bibitem[Feigelson \& Montmerle(1999)]{fei2} Feigelson, E. D., Montmerle, T., 1999, ARA\&A, 37, 363-408.

\bibitem[Feigelson \& Kriss(1981)]{fei3} Feigelson, E. D.; Kriss, G. A., 1981, ApJ, 248, L35-L38.

\bibitem[Flaherty et al.(2007)]{fla} Flaherty, K.M., Pipher, J, L, Megeath, S.T., et al., 2007, ApJ, 663, 1069F.

\bibitem[Freeman et al.(2002)]{fre} Freeman, P.~E., Kashyap, V., Rosner, R., \& Lamb, D.~Q.\ 2002, \apjs, 138, 185 

\bibitem[Frogel \& Persson(1974)]{fro} Frogel, J.A., Persson, S.E., 1974, ApJ, 192, 351.    

\bibitem[Getman et al.(2005)]{get} Getman, K. V., Flaccomio E., Broos, P.S., et al., 2005, ApJS, 160, 319-352.

\bibitem[Getman et al.(2006)]{get06} Getman, K. V., Feigelson, E. D., Townsley, L., et al., 2006, ApJS, 163, 306.

\bibitem[Giampapa et al.(1996)]{giam} Giampapa, M.S., Rosner, R., Kashyap, V., et al., 1996, ApJ, 463, 707. 

\bibitem[Gordon et al.(2005)]{gor} Gordon, K. D., Rieke, G.H., Engelbracht, C.W., et al. 2005, PASP, 117, 503.

\bibitem[Greene et al.(1994)]{gre} Greene, T.P., Wilking, B.A., Andre, P., et al., 1994, ApJ, 434, 614.

\bibitem[Gum(1955)]{gum} Gum, C.S., 1955, MNRAS, 67, 155.

\bibitem[Gutermuth et al.(2009)]{gut09} Gutermuth, R., Megeath, S.T., Myers, P.C., et al., 2009, ApJS, 184, 18G.  

\bibitem[Gutermuth et al.(2008b)]{gut07} Gutermuth, R., Myers, P.C., Megeath, S.T., et al., 2008, ApJ, 674, 336.

\bibitem[Gutermuth(2005)]{gut2} Gutermuth, R., 2005, PhD Thesis, University of Rochester.

\bibitem[Gutermuth et al.(2005)]{gut05} Gutermuth, R., Megeath, S. T., Pipher, J., et al., 2005, ApJ, 632, 397-420.  

\bibitem[Gutermuth et al.(2004)]{gut1} Gutermuth, R., Megeath, S. T., Muzerolle, J., et al.,  2004, ApJS, 154, 374.

\bibitem[Haisch et al.(2001)]{hai} Haisch, K. E. Jr., Lada, E. A. \& Lada, C. J. 2001, ApJ, 553, L153.

\bibitem[Harvey et al.(2006)]{har} Harvey, P., Chapman N., Lai S-P., 2005, ApJ, 644, 30-325.

\bibitem[Hayashi et al.(1996)]{hay} Hayashi, M.R., Hibata, K., Matsumoto. R., 1996, ApJ, 468, L37.

\bibitem[Hernadez et al.(2006)]{her} Hernadez, J., Briceno, C., Calvet, N., et al., 2006, ApJ, 652, 472.  

\bibitem[Hernadez et al.(2007)]{her2} Hernadez, J., Clavet, N., Briceno, C., et al., 2007, ApJ, 662, 1067.  

\bibitem[Isobe et al.(2003)]{iso} Isobe, H., Shibata, K., Yokoyama, T., Imanishi, K., 2003, PASJ, 55, 967.

\bibitem[Jorgensen et al.(2006)]{jor} Jorgensen, J., Harvey, P., Evans, N.J., et al, 2006, ApJ, 645, 1246-1263.

\bibitem[Kastner et al.(2002)]{kas} Kastner, J.H., Huenemoerder, D.P., Schulz, N.S., et al., 2002, ApJ, 567, 434.

\bibitem[Kenyon et al.(1993)]{ken} Kenyon, S. J., Calvet, N., Hartmann, L., 1993, ApJ, 414, 676.

\bibitem[Knodlseder(2000)]{knod} Knodlseder, J., 2000, A\&A, 360, 539.   

\bibitem[K\"{u}ker \& R\"{u}diger (1999)]{kuk} K\"{u}ker, M., \& R\"{u}diger, G., 1999, A\&A, 346, 922.

\bibitem[Lada \& Wilking(1984)]{lad84} Lada, C.J., Wilking. B.A., 1984, ApJ, 287, 610-621.

\bibitem[Lada(1987)]{lad} Lada, C.J., IAU Symp. 115, 1987, Star Forming Regions ed. M Peimbert \& J. Jugaku.

\bibitem[Lada et al.(2006)]{lad06} Lada, C.J., Luhman, K., Muench, A., et al., 2006, AJ, 131, 1574.

\bibitem[Landsman(1989)]{lan} Landsman, W.B., 1989, BAAS, Vol. 21, p784.

\bibitem[Megeath et al.(2004)]{meg} Megeath, S.T., Gutermuth R.A., Allen L.E., et al., 2004, ApJS, 154, 367.

\bibitem[Morrison \& McCammon(1983)]{mor} Morrison, R., McCammon, D., 1983, ApJ, 270, 119-122.

\bibitem[Muench et al.(2008)]{mue} Muench, A., Getman, K., Hillenbrand, L., Preibisch, T., 2008, in "Handbook of Star Forming Regions Vol.I", ASP, editor: Reipurth, B.. 

\bibitem[Preibisch \& Feigelson(2005)]{pre3} Preibisch, T., Feigelson, E., 2005, ApJS, 160, 390.  

\bibitem[Reach et al.(2005)]{rea} Reach, W.T., Megeath, S.T., Cohen, M., et al., 2005, PASP, 117, 978-990.

\bibitem[Rieke et al.(2004)]{rie} Rieke, G.H., Young, E.T., Engelbracht, C.W., et al., 2004, ApJS, 154, 25-29.

\bibitem[Rodgers, Campbell, Whiteoak(1960)]{rcw} Rodgers, A.W., Campbell, C.T., Whiteoak, J.B., 1960, MNRAS, 121, 103.  

\bibitem[Romanova et al.(2004)]{rom} Romanova, M.M., Ustyugova, G.V., Koldoba, A.V., Lovelace, R.V.E, 2004, ApJ, 616, L151.

\bibitem[Ryter(1996)]{ryt}  Ryter, Ch. E., 1996, 1996, Ap\&SS, 236, 285-291.   

\bibitem[Siess et al.(2000)]{sie} Siess, L., Dufour, E., Forestini, M., 2000, A\&A, 358, 593-599.  

\bibitem[Skrutskie et al.(2006)]{skr}  Skrutskie, M.F., Cutri R.M., Stiening R., et al., 2006, AJ, 131, 1163.

\bibitem[Smith et al.(1999)]{smi} Smith, C.H., Bourke, T.L., Wright, C.M., et al., 1999, MNRAS, 303, 367-379.   

\bibitem[Smith et al.(2001)]{smith}  Smith, R.K., Brickhouse, N.S., Liedahl, D.A., Raymond, J.C.,  2001, ApJ, 556, L91-L95.      

\bibitem[Stepnik et al.(2003)]{step} Stepnik, B., Abergel, A., Bernard, J-P, et al. 2003, A\&A, 398, 551.  

\bibitem[Stern et al.(1994)]{ste} Stern, R.A., Schmitt, J.H.M.M., Pye, J.P., et al., 1994, ApJ, 427, 808.  

\bibitem[Telleschi et. al.(2007)]{tell}  Telleschi, A., Gudel, M., Briggs, K.R., Audard, M., Palla, F., 2007, A\&A, 468, 425.  

\bibitem[van de Bergh \& Herbst(1975)]{van} van deBergh, S., Herbst, W., 1975, AJ, 80, 208.  

\bibitem[Vigil(2004)]{vig} Vigil M., 2004, M.Sc. Thesis, Massachusetts Institute of Technology.

\bibitem[Vuong et al.(2003)]{vuo} Vuong, M. H., Montmerle, T., Grosso, N., et al., 2003, A\&A, 408, 581-599.

\bibitem[Wang et al.(2008)]{wan} Wang, J., Townsley, L.K., Feigelson, E.D., et al., 2008, ApJ, 675, 464-490.  

\bibitem[Wang et al.(2011)]{wan11} Wang, J.,  Feigelson, E.D., Townsley, L.K., et al., 2011, arXiv1103.0785W.  

\bibitem[Whitney et al.(2003a)]{whi1} Whitney, B. A., Wood, K., Bjorkman, J. E., Wolff, M. J., 2003, ApJ, 591, 1049-1063.

\bibitem[Whitney et al.(2003b)]{whi2} Whitney, B. A., Wood, K., Bjorkman, J. E., Cohen, M., 2003, ApJ, 598, 1079-1099.

\bibitem[Winston et al.(2007)]{winston} Winston, E., Megeath, S.T., Wolk, S.J., et al.,  2007, ApJ, 669, 493.

\bibitem[Winston et al.(2009)]{win09} Winston, E., Megeath, S.T., Wolk, S.J., et al. 2009, AJ, 137, 4777.

\bibitem[Winston et al.(2010)]{win10} Winston, E., Megeath, S.T., Wolk, S.J., et al., 2010, AJ, 140, 266.

\bibitem[Wolk et al.(2006)]{wol}   Wolk, S. J., Spitzbart, B. D., Bourke, T. L., Alves, J., 2006, AJ, 132, 1100-1125. 

\bibitem[Wolk, Bourke \& Vigil(2008)]{wolkhandbook}   Wolk, S. J., Bourke, T. L., Vigil, M., 2008, in "Handbook of Star Forming Regions Vol.II", ASP, editor: Reipurth, B.. 

\bibitem[Wolk et al.(2008)]{wol2}   Wolk, S. J., Spitzbart, B., Bourke, T.L., et al., 2008, AJ, 135, 693.

\bibitem[Wolk et al.(2010)]{wol3}  Wolk, S. J., Winston, E., Bourke, T. L., et al., 2010, ApJ, 715, 671. 

\bibitem[Wolk et al.(2011)]{wol11} Wolk, S.J., Broos, P.S., Getman, K.V., et al., 2011, arXiv1103.1126W.

\bibitem[Yamaguchi et al.(1999)]{yama}  Yamaguchi, R., Saito, H., Mizuno, N., et al., 1999, PASJ, 51, 791.

\end{thebibliography}
\end{document}